\documentclass{aa}
\usepackage{graphicx,amsmath}
\usepackage{txfonts}

\usepackage{natbib} 
\bibpunct{(}{)}{;}{a}{}{,}

\begin{document}
\title{Application of speckle and (multi-object) multi-frame blind deconvolution techniques on imaging and imaging spectropolarimetric data}

   \author{K.G.Puschmann\inst{1,2} \and C.Beck\inst{2}}
        
   \titlerunning{Application of speckle and MOMFBD/MFBD techniques on 2D data}
  \authorrunning{K.G.Puschmann \& C.Beck}  
   \institute{Leibniz-Institut f\"ur Astrophysik Potsdam (AIP)
     \and Instituto de Astrof\'{\i}sica de Canarias
    }
\date{Received xxx; accepted xxx}
\abstract{Ground-based imaging and imaging spectropolarimetric data are often subjected to post-facto reconstruction techniques to improve the spatial resolution.}
{We test the effects of reconstruction techniques on two-dimensional data to determine the best approach to improve our data.}{We obtained an 1-hour time-series of spectropolarimetric data in the \ion{Fe}{i} line at 630.25\,nm with the G{\"o}ttingen Fabry-P{\'e}rot Interferometer (FPI) that are accompanied by imaging data in the blue continuum at 431.3\,nm and \ion{Ca}{ii} H at 396.85\,nm. We apply both speckle and (multi-object) multi-frame blind deconvolution ((MO)MFBD) techniques. We use the ``G\"ottingen'' speckle and speckle deconvolution codes and the MOMFBD code in the implementation of Van Noort et al.~(2005). We compare the resulting spatial resolution and investigate the impact of the image reconstruction on spectral characteristics of the G\"ottingen FPI data.}
{The speckle reconstruction and MFBD perform similar for our imaging data with nearly identical intensity contrasts. MFBD provides a better and
  more homogeneous spatial resolution at the shortest wavelength when applied
  to a series of image bursts. The MOMFBD and speckle deconvolution of the
  intensity spectra lead to similar results, but our choice of settings for the MOMFBD yields an intensity contrast smaller by about 2\,\% at a comparable spatial
  resolution. None of the reconstruction techniques introduces significant artifacts in
  the intensity spectra. The speckle deconvolution (MOMFBD) has a rms noise in
  Stokes $V/I$ of 0.32\,\% (0.20\,\%). The
  deconvolved spectra thus require a high significance threshold of about 1.0\,\%  to separate noise peaks
  from true signal. A comparison to spectra with a significantly higher
  signal-to-noise (S/N) ratio and to spectra from a magneto-hydrodynamical simulation reveals that the G{\"o}ttingen FPI can only detect about 30\,\% of the polarization signal present in quiet Sun areas. The distribution of NCP values for the speckle-deconvolved data matches that of observations with higher S/N better than MOMFBD, but shows seemingly artificially sharp boundaries and unexpected changes of the sign.}{For our imaging data, both MFBD and speckle reconstruction are equivalent, with a slightly better and more stable performance of MFBD. For the spectropolarimetric data, the higher intensity contrast of the speckle deconvolution is balanced by the smaller amplification of the noise level in the MOMFBD at a comparable spatial resolution. The noise level prevents the detection of weak and diffuse magnetic fields. Future efforts should be directed to improve the S/N of the G{\"o}ttingen  FPI spectra for spectropolarimetric observations to lower the final significance thresholds.}
\keywords{Sun: photosphere -- Sun: spectropolarimeters}
 \maketitle
\section{Introduction}
One of the primary objectives of modern solar physics is the determination of
the thermodynamic and magnetic structure of the solar atmosphere at the smallest spatial scales, including isolated magnetic elements in the quiet Sun and the internal fine-structure of larger features such as sunspots. An accurate derivation of the atmospheric properties requires both high spatial and spectral resolution. Telescopes with apertures of 1\,--\,1.5\,m, such as the Swedish Solar Telescope \citep[SST,][]{scharmer+etal2003}, GREGOR \citep{volkmeretal07,balthasar+etal2007}, or the New Solar Telescope \citep[][]{denkeretal06},
allow one to study the dynamics of the solar fine-structure at spatial scales down to 50\,--\,100 km on the surface of the Sun. The next-generation solar telescopes of the 4-m class, i.e., the Advanced Technology Solar Telescope \citep [][]{wagneretal08} or the European Solar Telescope \citep [][]{collados08}, then will resolve the fundamental scale of the solar photosphere, i.e., the photon free path length.

All observations from the ground are, however, degraded by the wavefront distortions caused by fluctuations of the refractive index in the Earth's atmosphere. To overcome this limitation, adaptive optics (AO) systems are used to correct the wavefront errors, but their compensation is only partial and the diffraction limit of the respective telescope is hard to reach. AO systems have been installed at the Dunn Solar Telescope \citep[DST,][]{rimmele+radick98, rimmele2004a}, the German Vacuum Tower Telescope \citep[VTT,][]{vdluehe+etal2003}, and the SST \citep{scharmer+etal2003a}. AO and multi-conjugated AO systems \citep{beckers1988,berkefeld+etal2001,moretto+etal2004,ellerbroek+vogel2009} with a larger corrected field of view (FOV) are also foreseen for all next-generation telescopes.

For spectroscopic or spectropolarimetric observations taken with slit-spectrograph systems, it is difficult to improve the spatial resolution beyond the limit provided by the real-time correction of an AO system because of the essentially one-dimensional information at one position of the slit \citep[see, e.g.,][submitted to A\&A]{keller+johannesson1995,suetterlin+wiehr2000,aschwanden2010,beck+rezaei2011}. Contrary to that, instruments providing instantly a two-dimensional FOV offer the possibility for a post-facto reconstruction of the observations. Two-dimensional (2D) spectrometers are commonly realized in the form of Fabry-P{\'e}rot Interferometers (FPIs) that have a narrow tunable spectral band-pass and a high transmission.

A first example of a 2D FPI spectrometer with three etalons was designed for the Australian solar facility in Narrabri \citep{ramsayetal70, loughheadetal78, bray88}. Nowadays, almost all ground-based solar telescopes are equipped with FPI instruments: the Italian Panoramic Monochromator \citep{bonaccinietal89,cavallini98} at the THEMIS telescope, the Interferometric BIdimensional Spectrometer \citep[IBIS,][]{cavallinietal00, cavallini2006, reardon+cavallini2008} at the DST, the visible-light and near-infrared imaging magnetographs \citep{denkeretal03} at the Big Bear Solar Observatory, the CRisp Imaging Spectro Polarimeter \citep[CRISP, e.g.,][]{vannoort+vandervoort2008,scharmeretal08} at the SST, and the Triple Etalon SOlar Spectrometer \citep[TESOS,][]{kentischeretal98, vdluehe+kentischer00, tritschleretal02} at the VTT. For the VTT, another 2D instrument was developed prior to TESOS by the Institute for Astrophysics G\"ottingen, the so-called G\"ottingen FPI \citep[][]{bendlinetal92, bendlin93, bendlin+volkmer93, bendlin+volkmer95}. This instrument will in the future be one of the post-focus instruments at the GREGOR telescope. Therefore, the spectrometer was largely renewed during the first half of 2005 \citep{puschmann+etal2006}. Its design as the new ``Gregor FPI'' is described in \citet{puschmannetal07} and \citet{denker+etal2010}.

Even if most of the 2D instruments were initially designed as spectrometers
only, they now have all been upgraded for (vector) spectropolarimetry
(see e.g., \citet{nazi+kneer2008a} and \citet{balthasaretal09} for the G{\"o}ttingen FPI;
\citet{viticchi+etal2009} and \citet{judge+etal2010} for IBIS, and
\citet{beck+etal2010} for TESOS). All 2D instruments use at least two
channels, where one beam passes through the tunable FPIs with their high spectral resolution (narrow-band (NB) channel). The second channel uses only a broad-band (BB) interference filter to provide strictly simultaneous images with a high signal-to-noise (S/N) ratio for an accurate alignment of the individual wavelength steps and the application of post-facto reconstruction techniques.

 Two main reconstruction techniques have been developed and successfully applied to observation data up to now: the speckle reconstruction technique with the possibility of a subsequent NB deconvolution of spectroscopic or spectropolarimetric data, and the blind deconvolution (BD) technique that can be applied to both imaging and 2D spectropolarimetric data.

For a speckle reconstruction, a fast series of about 50\,--\,100 short-exposed images $I_{\rm obs}({\bf x},t_i)$ is taken at times $t_i$. The observed images represent the image of the real object $I_{\rm real}({\bf x})$ modulated by the instantaneous optical transfer function (OTF) of the Earth's atmosphere and the telescope. In Fourier space, the corresponding equation reads 
\begin{equation}
\tilde I_{\rm obs}({\bf s}, t_i) = \tilde I_{\rm real}({\bf s}) \cdot {\rm OTF} ({\bf s}, t_i) \;, \label{eq_otf}
\end{equation}
where $^\sim$ denotes the Fourier transform, and ${\bf x}$ and ${\bf s}$ are the spatial coordinates and their corresponding Fourier equivalent, respectively.

To recover the information of the real image, the Fourier amplitude and phases of $\tilde I_{\rm real}$  are derived separately. The Fourier amplitudes are usually determined by the method of \citet{labeyrie1970} and the spectral ratio method \citep[][]{vdluehe1984}. Two main approaches exist for the estimate of the Fourier phases, i.e., the Knox-Thompson and speckle masking algorithm \citep{knox+thompson1974,weigelt1977}. The large number of input images is necessary to obtain good atmospheric statistics. For the application to solar observations with low-contrast images, several specific speckle reconstruction methods were developed \citep[see, e.g.,][]{vdluehe1993,deboer93,mikurda+vdl2006}. Examples of recent speckle codes are the Kiepenheuer-Institute Speckle Interferometry Package \citep[KISIP,][]{woeger+etal2008,woeger+vdluehe2008} and the G\"ottingen speckle reconstruction code \citep{deboer93}. The latter was improved by \citet{puschmann+sailer2006} to account for the field-dependent real-time correction by AO (see also \citet{denker+etal2007}, or \citet{woeger2010} for the KISIP code) and successfully applied in, e.g., \citet{puschmannwiehr06}, \citet{blancoteal07}, or \citet{sanchezetal08}. For 2D spectro(polari)meters, the BB images can be directly speckle reconstructed because a sufficient number of images is available. The simultaneously recorded NB images -- always only a few images per wavelength position -- can then be deconvolved using the method of \citet{keller+vdluehe1992} estimating the instantaneous OTF from the observed and reconstructed BB images \citep[see also][]{kriegetal99,janssen03,bellogonzalezetal05}.

In the case of the BD technique, one estimates both the real image {\em and} the OTF at the same time without using an OTF model based on the seeing statistics. This leads to a problem similar to the disentanglement of magnetic
field strength and area filling factor for spectral lines in the weak-field
limit: an infinite number of possible combinations can yield nearly the same
result. The ambiguity is lessened by the use of multiple image frames, if the
real image is assumed to be always the same during the observed fast image
sequence. Compared with the speckle reconstruction, a significantly smaller
number of images, about 5\,--\,20, is sufficient \citep[see,
e.g.,][]{loefdahl+etal2007}. Such a multi-frame blind deconvolution
\citep[MFBD,][]{schulz93,vankampen+paxman98,loefdahl2002} can be applied to
series of images of a single real object. The approach was extended by \citet{loefdahl2002} and \citet{vannortetal05} to cover also the case of multiple objects, e.g., simultaneous images of the same object in slightly different wavelengths or different polarization states. This allows the reconstruction of the simultaneously recorded BB and spectro(polari)metric NB
images of, for instance, FPI instruments. The corresponding multi-object multi-frame blind deconvolution \citep[MOMFBD,][]{vannortetal05} code is freely available\footnote{http://www.momfbd.org/}. A drawback of the (MO)MFBD is the necessary computing effort that usually requires the use of a dedicated computer cluster or similar installations, whereas speckle codes can also be run within limitations on a single desktop PC or nearly in real-time on a computer cluster \citep{denker+etal2001}.

Even if the application of reconstruction methods to imaging and 2D
spectroscopic or spectropolarimetric data is a routine task these days, there
are only relatively few investigations on the influence of the reconstruction
on the data properties, e.g., \citet{mikurda+etal2006} for
spectroscopic observations or \citet{vannoort+vandervoort2008} for
MOMFBD-reconstructed data. The reconstruction introduces more severe changes of
the data than, e.g., a simple flat fielding, and therefore also may introduce systematic artifacts. 

FPI instruments sequentially scan the wavelength points; the {\em independent} observation and reconstruction of individual wavelength steps thus may compromise the spectral quality. This is even worse for spectropolarimetric observations, where additionally the information on the polarization signal is encoded in the intensity of various modulation states. An independent scaling of the intensity in the individual images -- varying across the FOV as well -- therefore may easily introduce spurious polarization signals because they are finally derived from intensity differences. 

In this contribution, we investigate the performance of speckle and BD
reconstruction techniques. Because our observations provided us with both
imaging and 2D spectropolarimetric data, we extended the study to both types
of data. In Sect.~\ref{sect_obs}, we describe the observational setup and the
data acquisition. Section \ref{sect_datared} explains the data reduction and
reconstruction methods for the individual data sets. The results presented in
Sect.~\ref{sect_results} are summarized and discussed in
Sect.~\ref{sect_summ}. Section \ref{concl} provides the conclusions.
\section{Observations\label{sect_obs}}
The observations consist of a time-series of 65 minutes taken on 2009 June 7 between 07:45 and 08:50 UT at the VTT with real-time correction of seeing distortions by the Kiepenheuer-Institute AO system \citep[KAOS,][]{vdluehe+etal2003}. The Fried parameter $r_{\rm0}$ as estimated by the KAOS system was on average about 10\,cm during the observations. The target area was a quiet Sun (QS) region at disk center. 
\subsection{Setup}
\begin{figure}
\centerline{\resizebox{7cm}{!}{\includegraphics{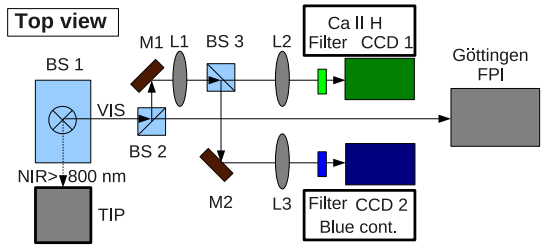}}}
\caption{Setup of the observations in a top view. BS1: dichroic, splitting
  into NIR ($>$\,800\,nm) and VIS light. BS2: dichroic with its edge at 450\,nm, splitting between the G{\"o}ttingen FPI and the imaging channels. BS3: achromatic, 50/50. The two imaging channels in \ion{Ca}{ii} H and blue continuum used BB prefilters in front of the respective CCD cameras.\label{fig_setup}}
\end{figure}
We used the setup prepared for the ground-based support observations during the Sunrise flight (see Fig.~\ref{fig_setup}), following the concept for multi-wavelength observations at the VTT described in \citet{beck+etal2007a}. The light from the telescope was first split into its visible (VIS) and near-infrared (NIR) parts by a dichroic beamsplitter (BS1) with a separation edge at 800\,nm. The NIR light ($>$\,800\,nm) was transmitted towards the Echelle spectrograph, where the Tenerife Infrared Polarimeter \citep[TIP,][]{martinez+etal1999, collados+etal2007} was mounted. TIP took high cadence spectra of the 1083\,nm range at a fixed slit position. These data, however, will not be used in the current study. The VIS part was reflected towards the G{\"o}ttingen FPI. Two imaging channels in blue continuum (BC) and \ion{Ca}{ii} H were fed from the VIS beam by a dichroic pentaprism (BS2) that reflects all light below 450\,nm to the side. The reflected light was evenly shared between the two imaging channels using an achromatic 50/50 beamsplitter (BS3). For the final wavelength selection, BB interference filters (431.3$\pm 0.3$\,nm and 396.85$\pm$0.3\,nm) were mounted directly in front of the two imaging cameras. 
\subsection{Data properties}
\paragraph{Imaging data} The two imaging channels used PCO 4000 cameras in a 2$\times$2 binning mode yielding a spatial sampling of 0\farcs071 per pixel for the blue continuum channel and 0\farcs076 per pixel for \ion{Ca}{ii} H. The FOV was about $110^{\prime\prime}\times 80^{\prime\prime}$. The cameras were run independently of the rest of the instruments in a streaming mode. Because the disk write time was different due to the image size and the intrinsic write time of the PC hard disks, we added a short delay time to the faster channel to obtain nearly identical frame rates. We then finally obtained bursts of 100 (96) images each 17 seconds with an exposure time of 7.5 (20) ms for the blue continuum channel (\ion{Ca}{ii} H). During the first 50 min of uninterrupted observations, we collected 200 bursts in each channel that were individually reconstructed.
\paragraph{G{\"o}ttingen FPI data} The vector-polarimetric mode of the G{\"o}ttingen FPI in its latest setup at the VTT was used \citep{puschmann+etal2006,nazi+kneer2008a}. We scanned the \ion{Fe}{i} line at 630.25\,nm in 28 steps of 2.31 pm step width in wavelength, covering a range from about 630.23\,nm to 630.29\,nm. At each wavelength step, four modulation states were acquired with an exposure time of 20\,ms for each image. The modulation sequence was then repeated eight times to have a sufficient number of images for the later deconvolution. Strictly simultaneous BB images (28$\times$4$\times$8\,=\,896 images per wavelength scan) were taken on a second camera with a BB prefilter centered at 630\,nm. We chose a frame rate of 10\,Hz for the observations because all higher frame rates led sporadically to a complete loss of the synchronization between the modulation by the liquid crystals and the camera read-out. The cadence for a line scan was therefore limited to 90 seconds. The spatial sampling was 0\farcs11 per pixel across a FOV of $38^{\prime\prime}\times 58^{\prime\prime}$ before the reconstruction. The final size of the FOV then slightly varies depending on the reconstruction method. 
\subsection{Additional data \label{add_data}}
For comparison with the spectra from the G\"ottingen FPI, we will use three additional data sets. One is an observation taken with the Hinode spectropolarimeter \citep[SP,][]{kosugi+etal2007} that is described in more detail in \citet{puschmann+etal2010,puschmann+etal2010a}. This scan with the Hinode/SP was taken on the 1$^{st}$ of May 2007 and used an integration time of 4.8\,s per scan step of about 0\farcs15 width, covering an active region close to disk center. The root-mean-square (rms) noise level in Stokes $V$ in a continuum wavelength window was about $1.2\times 10^{-3}$ of $I_c$. We cut out a $20^{\prime\prime}\times 20^{\prime\prime}$ QS section of its FOV for this study. The second observation used was taken with the POlarimetric LIttrow Spectrograph \citep[POLIS,][]{beck+etal2005b}. We selected an observation of QS on disk center obtained on the 29$^{th}$ of Aug 2009, cutting out again a $20^{\prime\prime}\times 20^{\prime\prime}$ section. The integration time per scan step of 0\farcs5 width was 26\,secs, yielding a rms noise of about $3\times 10^{-4}$ of $I_c$ (see the POLIS data archive\footnote{http://www3.kis.uni-freiburg.de/$\sim$cbeck/POLIS\_archive/\\POLIS\_archivemain.html} for more details). These data provide the best reference for the polarization signals to be expected in QS because of their low noise level. As third data set, we used spectra derived from a magneto-hydrodynamic (MHD) simulation. The MHD simulation was performed with the CO$^5$BOLD code \citep{freytag+etal2002} and is described in more detail in \citet{schaffenberger+etal2005,schaffenberger+etal2006}. For one simulation snap shot, spectra of the two \ion{Fe}{i} lines at 630\,nm were synthesized with a spectral sampling of 2\,pm per pixel \citep[for more details see][]{steiner+etal2008,beck+rammacher2010}. Figure \ref{ncp_fig} later on shows maps of the continuum intensity $I_c$ for the three additional data sets.
\section{Data reduction\label{sect_datared}}
\subsection{Imaging data} The data from the two imaging channels (and the G{\"o}ttingen FPI) were corrected for flat field inhomogeneities and the dark current. For the imaging data, we then applied two reconstruction methods to every burst of about 100 images: the modified version of the ``G{\"o}ttingen'' speckle code \citep{puschmann+sailer2006} and the MFBD method. For the latter, we used the version of the algorithm of \citet{vannortetal05} installed at the Instituto de Astrof\'isica de Canarias (IAC) that is prepared for parallel computing. 
\subsection{G{\"o}ttingen FPI data} The spectra of the G{\"o}ttingen FPI consist of a set of NB intensity images $I(\lambda_i , m_j , k)$, where the index $i=1,\hdots,28$ denotes the number of wavelength steps, $m_j$ the four modulation states $j=1,\hdots,4$, and $k=1,\hdots,8$ the index of repetitions of the modulation cycle at one wavelength step. A simultaneous set of BB images  $B(\lambda_i , m_j , k)$ is available from the second camera of the G{\"o}ttingen FPI. The two orthogonally polarized beams in the NB images ($I^+$ and $I^-$) and the corresponding FOV of the BB images were extracted and aligned with pixel accuracy to avoid an interpolation of the data prior to any image reconstruction. The G{\"o}ttingen FPI spectra were then reduced with three different methods: simple destretching, speckle deconvolution, and MOMFBD. 
\paragraph{Destretching} For destretching and later also the speckle deconvolution, we first speckle-reconstructed the 896 BB images of each wavelength scan to obtain a reference image $B_{\rm speckle}$. For destretching, we then determined the necessary shifts in the FOV to co-register the individual BB images during the wavelength scan with the speckle reference image and applied the same corrections to the NB images $I^\pm(\lambda_i , m_j , k)$. We then averaged the co-registered eight repetitions of each modulation state to get $I^\pm(\lambda_i , m_j) = 1/8 \sum_{k=1}^8 I^\pm(\lambda_i , m_j , k)$. Destretching and speckle deconvolution were applied separately to the two beams $I^+$ and $I^-$, whereas the MOMFBD deconvolves both beams at the same time (see Eq.~(\ref{momfbd_eq}) below).
\paragraph{Speckle deconvolution} For the reconstruction of the spectropolarimetric data we adapted the G\"ottingen NB deconvolution code \citep{kriegetal99,janssen03} for an application to vector polarimetric measurements. The eight individual NB images $I(\lambda_i, m_j,k)$ of each modulation state $m_j$ and each wavelength $\lambda_i$ are deconvolved using the information obtained from the simultaneous BB images and the speckle-reconstructed reference image $B_{\rm speckle}$. In the Fourier domain, the corresponding equation reads
\begin{equation}
\tilde I^\pm (\lambda_i , m_j) = F \, \frac{ \sum_{k=1}^{8} \tilde I^\pm (\lambda_i, m_j,k) \tilde B^*(\lambda_i, m_j,k)}{\sum_{k=1}^{8} |\tilde B (\lambda_i, m_j,k)|^2 } \, \tilde B_{\rm speckle} \;, \label{eq_speckle}
\end{equation}
where $^\sim$ denotes the Fourier transform, $^*$ the complex conjugate, and $F$ is a noise filter \citep[optimum filter, see][]{keller+vdluehe1992}. The inverse Fourier transform then yields the wanted $I^\pm (\lambda_i , m_j)$. The deconvolution is applied individually to overlapping subfields of 64$\times$64 pixels because Eq.~(\ref{eq_speckle}) is only valid inside an isoplanatic patch \citep{vdluehe1993}.  
\paragraph{MOMFBD} For the MOMFBD, all images from a single wavelength scan were fed simultaneously into the code to obtain one reconstructed BB image $B_{\rm real}$ and one image per modulation state, wavelength position, and orthogonally polarized beam, $I^\pm(\lambda_i , m_j)$. We used the default settings of the MOMFBD code which employs a decomposition of the wavefront phases into 36 Karhunen-Lo{\`e}ve modes and an equal weight for NB and BB data. The code then minimizes 
\begin{eqnarray}
L&&= \sum_{i = 0}^{28}\sum_{j=0}^4  \sum_{k=1}^8 \sum_{\bf S} | \tilde I_{\rm obs}^+(\lambda_i, m_j,k) - \tilde I_{\rm real}^+(\lambda_i, m_j) \cdot {\rm OTF}(i,j,k)|^2 ({\bf s}) \nonumber \\
&&+| \tilde I_{\rm obs}^-(\lambda_i, m_j,k) - \tilde I_{\rm real}^-(\lambda_i, m_j) \cdot {\rm OTF}(i,j,k)|^2 ({\bf s})\nonumber \\ 
&&+| \tilde B_{\rm obs}(\lambda_i, m_j,k) - \tilde B_{\rm real} \cdot {\rm OTF}(i,j,k)|^2 ({\bf s})\,  \,, \label{momfbd_eq}
\end{eqnarray}
where ${\bf s}$ denotes the Fourier equivalent of the spatial coordinates \citep{loefdahl2002,vannortetal05}. This reconstruction is also performed for subfields of 64$\times$64 pixels. 

As became clear during the subsequent investigation of the deconvolved spectra, the solution with equal weight for the BB and NB channel in the MOMFBD is only one possible option with specific characteristics. We run a series of additional test deconvolutions that are described in Appendix \ref{appa}, but finally decided to stick to the initial settings because they provide the lowest noise level in Stokes $V$ at continuum wavelengths of all test runs and maintain the spectral quality best (see Sect.~\ref{motiv}).

All three reduction methods for G{\"o}ttingen FPI spectra yield finally a set of 28$\times$4 intensity images for the two beams, $I^\pm(\lambda_i , m_j)$. The spectra were then corrected for the transmission curve of the NB prefilter, using a separate wavelength scan obtained with a continuum light source. Afterwards, we applied a correction for the field-dependent blue shift in the spectra caused by the collimated mounting of the etalons. The necessary blue shift correction was determined from the line-core position of the solar \ion{Fe}{i} line in the average flat field frames. The polarization information could now be retrieved by the polarimetric calibration procedure described in the next section.
\subsection{Polarimetric calibration \label{sect_cal}}
The polarimetric calibration of the G{\"o}ttingen FPI data was carried out in a two-step procedure with a correction for the instrumental polarization induced by the telescope \citep{beck+etal2005a} and the cross-talk caused by the optics behind the instrumental calibration unit (ICU) of the VTT. The calibration data obtained with the ICU were used to determine the polarimeter response function with the generic approach for the VTT described in, e.g., \citet{beck+etal2005b} or \citet{beck+etal2010}. The efficiencies for measuring the components of the Stokes vector \citep[cf.][]{deltoroiniesta+collados2000} were about ($\epsilon_I,\epsilon_Q,\epsilon_U\epsilon_V) = (0.93,0.56,0.49,0.42)$ in our case, with a total efficiency for measuring polarization of $\epsilon_{\rm pol} = 0.85$. 

Residual $I\rightarrow QUV$ cross-talk was removed by forcing the continuum polarization level to zero. We determined the average ratio $\alpha_{QUV}=QUV/I$ in the continuum wavelength range from 630.275 to 630.291\,nm separately in each profile and then subtracted $\alpha_{QUV}\cdot I(\lambda)$ from Stokes $QUV(\lambda)$. This approach compensates automatically any possible variation of the cross-talk level across the FOV, but could only be used because continuum wavelength points without intrinsic polarization signal were available. Scanning only a smaller wavelength range around the solar spectral line does not permit to use this correction because it cannot be assumed that the polarization signal at these wavelengths should always be zero in real observations.
\begin{figure}
\centerline{\resizebox{8.8cm}{!}{\includegraphics{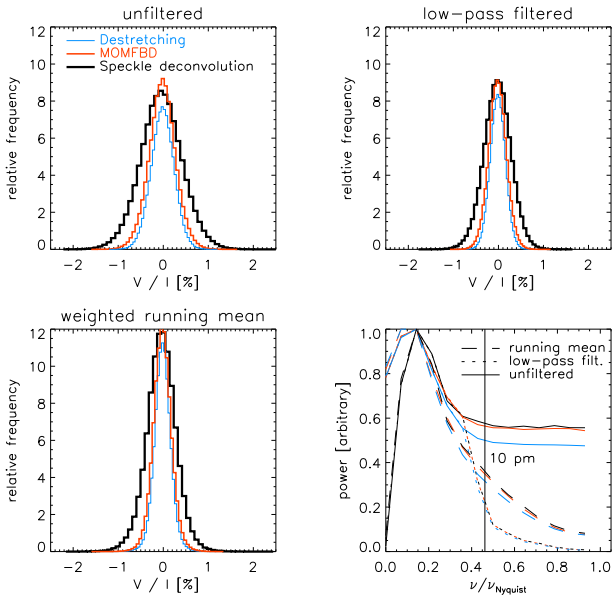}}}
\caption{Histograms of the polarization amplitude of Stokes $V/I$ in a continuum window. {\em Top left}: unfiltered data for speckle deconvolution ({\em black}), MOMFBD ({\em red}), and destretching ({\em blue}). {\em Top right} and {\em bottom left}: the same for application of a low-pass filter and of a weighted running mean, respectively. {\em Bottom right}: Fourier power vs.~spectral frequency. {\em Solid/dotted/dashed lines}: unfiltered/low-pass filtered/running mean. The {\em vertical black line} corresponds to a wavelength of 10\,pm.\label{fig_hist}}
\end{figure}
\subsection{Spectral noise filtering\label{noise_filter}}
The reconstruction process changes the intensities at different spatial locations and at different wavelength points independently of each other. Because the polarization signal is derived from intensity differences between the four images belonging to the different modulation states, a relative change of intensity by the reconstruction process can produce spurious polarization signals. Imperfections in the spatial alignment of different modulation states can also have the same effect. The spectra of the speckle-deconvolved data show a large noise level in the polarization signal of the Stokes $V/I$ signal at continuum wavelengths (630.275\,nm to 630.291\,nm) with peak values larger than 1.5\,\% ({\em top left panel} of Fig.~\ref{fig_hist}), which requires some sort of filtering. The noise level in the MOMFBD or destretched data is lower than that of the speckle-deconvolved data by a factor of about 1.5; the MOMFBD data show a slightly higher noise level than the destretched data (first column of Table \ref{tab_noise}).
\begin{figure*}
\sidecaption
\begin{minipage}{12cm}
\fbox{\resizebox{11.5cm}{!}{\includegraphics{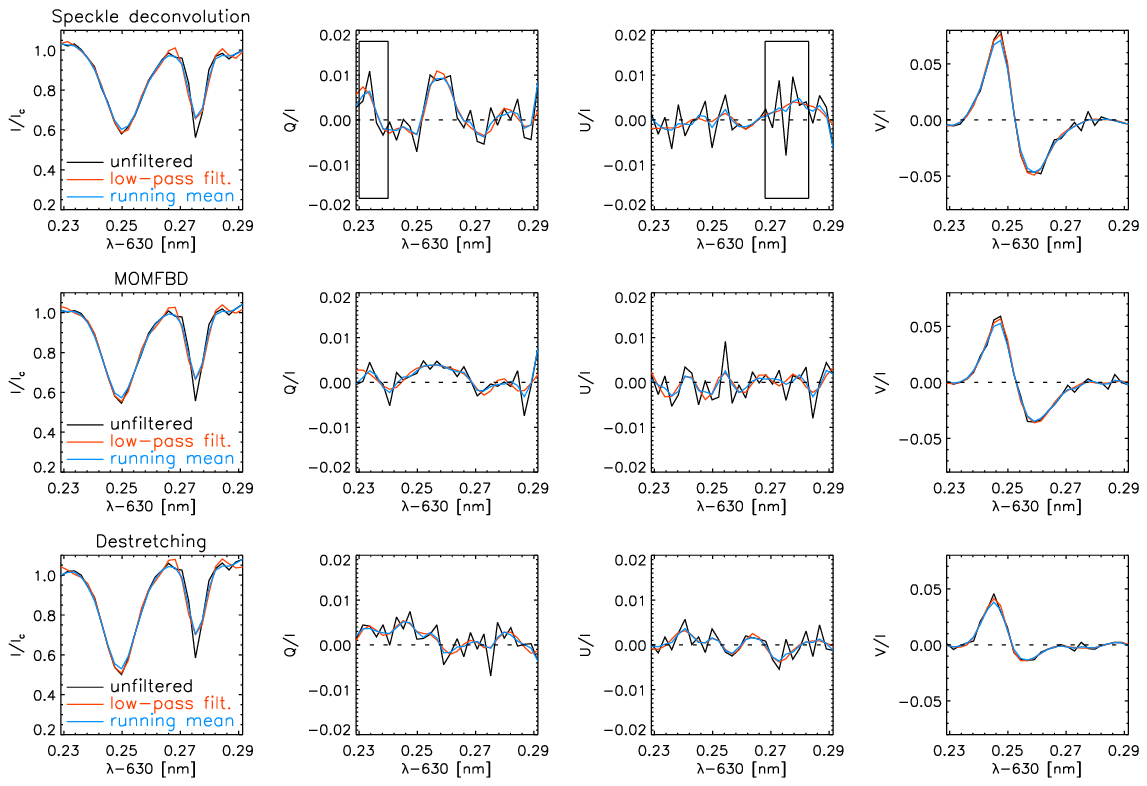}}}
\fbox{\resizebox{11.5cm}{!}{\includegraphics{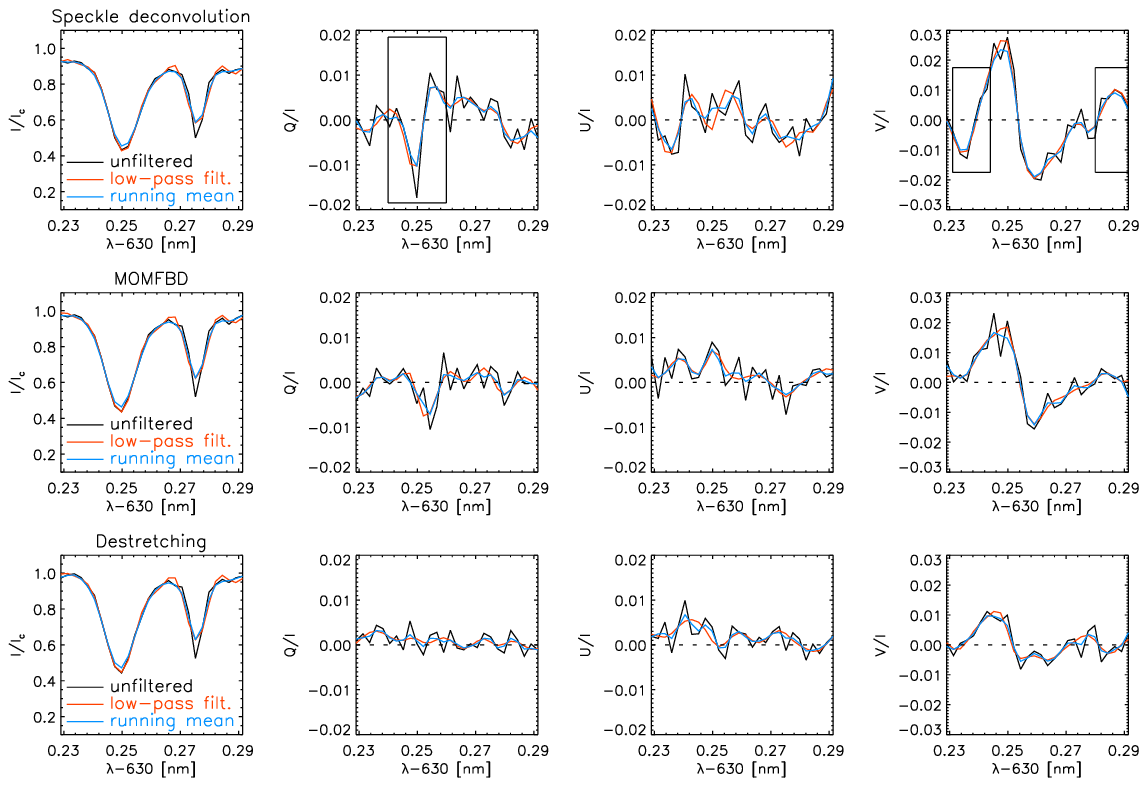}}}
\end{minipage}
\caption{Example spectra in the network ({\em top panel}) and in a small isolated polarization patch ({\em bottom panel}). {\em Left to right}: Stokes $IQUV$. {\em Top to bottom}: speckle deconvolution, MOMFBD, and destretched data. {\em Black/red/blue lines}: unfiltered data/low-pass filter/running mean. The {\em black rectangles} denote noise peaks that could be mistaken for solar polarization signal after the filtering. \label{fig_spec_filter}}
\end{figure*}

We used two different approaches for the spectral noise filtering: a low-pass filter as in \citet{nazi+etal2009} and a weighted running mean in the spectral dimension over three wavelength steps ($\equiv 6.9$\,pm). For the latter, we calculated the new values of Stokes $IQUV$ at the wavelength $\lambda_i$ by $\sum_{j=-1}^{j=+1} a_{j+1}\, IQUV(\lambda_{i+j})$ with {\bf a}= (0.25,0.5,0.25). The low-pass filter was set to cut all power for frequencies above half of the Nyquist frequency of $(2\times 2.31$\,pm$)^{-1}$. The {\em upper right} and {\em lower left} panels of Fig.~\ref{fig_hist} show the histograms of the Stokes $V$ continuum polarization signal after applying the two filters to the spectra. In both cases, the peak values of the noise reduce to about 1\% for the speckle deconvolution and to about 0.5\% for the other two data reduction methods. 
\begin{table}
\caption{Rms noise in Stokes $V/I$ in percent.\label{tab_noise}}
\begin{tabular}{cccc}\hline\hline
Type & unfiltered & filtered & running mean \cr\hline
Speckle deconvolution & 0.50 & 0.32 &  0.32 \cr
MOMFBD & 0.32 &  0.20 &  0.20 \cr
Destretching & 0.29 &   0.18 &  0.18 \cr\hline
\end{tabular}
\end{table}

In the {\em lower right} panel of Fig.~\ref{fig_hist}, the Fourier power as a function of the spectral frequency normalized with the Nyquist frequency is displayed \citep[cf.][their Fig.~1]{nazi+etal2009}. The power spectra show that the main power is concentrated at small frequencies corresponding to a large wavelength extent, but the power in the unfiltered data stays at about 50\% of the maximum value up to the frequency limit because of the high noise level. The weighted running mean has a very similar effect as the low-pass filter and reduces the power to about 20\% at half of the Nyquist frequency. For comparison, the full width at half maximum (FWHM) of the \ion{Fe}{i} line at 630.25\,nm of about 10 pm is marked with a {\em vertical black line}. Because any spectral features should be broadened to a comparable extent by thermal and Doppler broadening, none of the two filters should thus have removed significant solar information. 
\begin{figure*}
\centerline{\resizebox{15.9cm}{!}{\resizebox{13.9cm}{!}{\hspace*{.4cm}\includegraphics{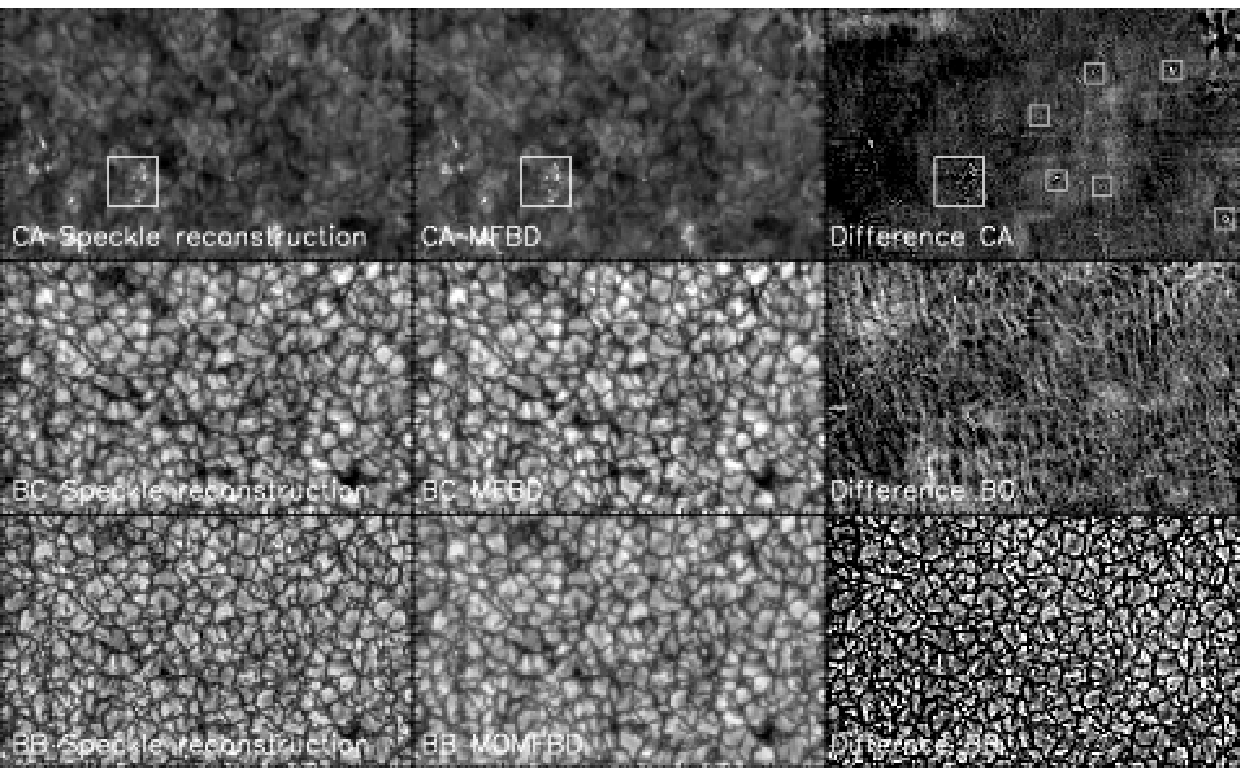}}\hspace*{.7cm}\resizebox{3.15cm}{!}{\hspace*{.3cm}\includegraphics{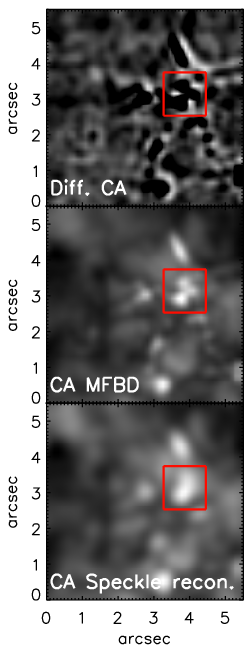}}}}$ $\\$ $\\
\caption{Overview maps of the imaging data. {\em Bottom to top}: G{\"o}ttingen FPI BB channel, blue continuum, and Ca. {\em Left to right}: speckle reconstruction, (MO)MFBD, and difference image. The display ranges for the intensity maps are $0.8<I<1.5$ for Ca and $0.8<I<1.2$ otherwise. The difference images are thresholded at $\pm 0.05$. The {\em large white rectangles} denote the region shown magnified {\em in the most right column}, the {\em small grey squares} indicate locations with ring structures in the Ca difference image. \label{fig_imaging_overview}}
\end{figure*}
\begin{figure*}
\centerline{\resizebox{15.9cm}{!}{\hspace*{.5cm}\includegraphics{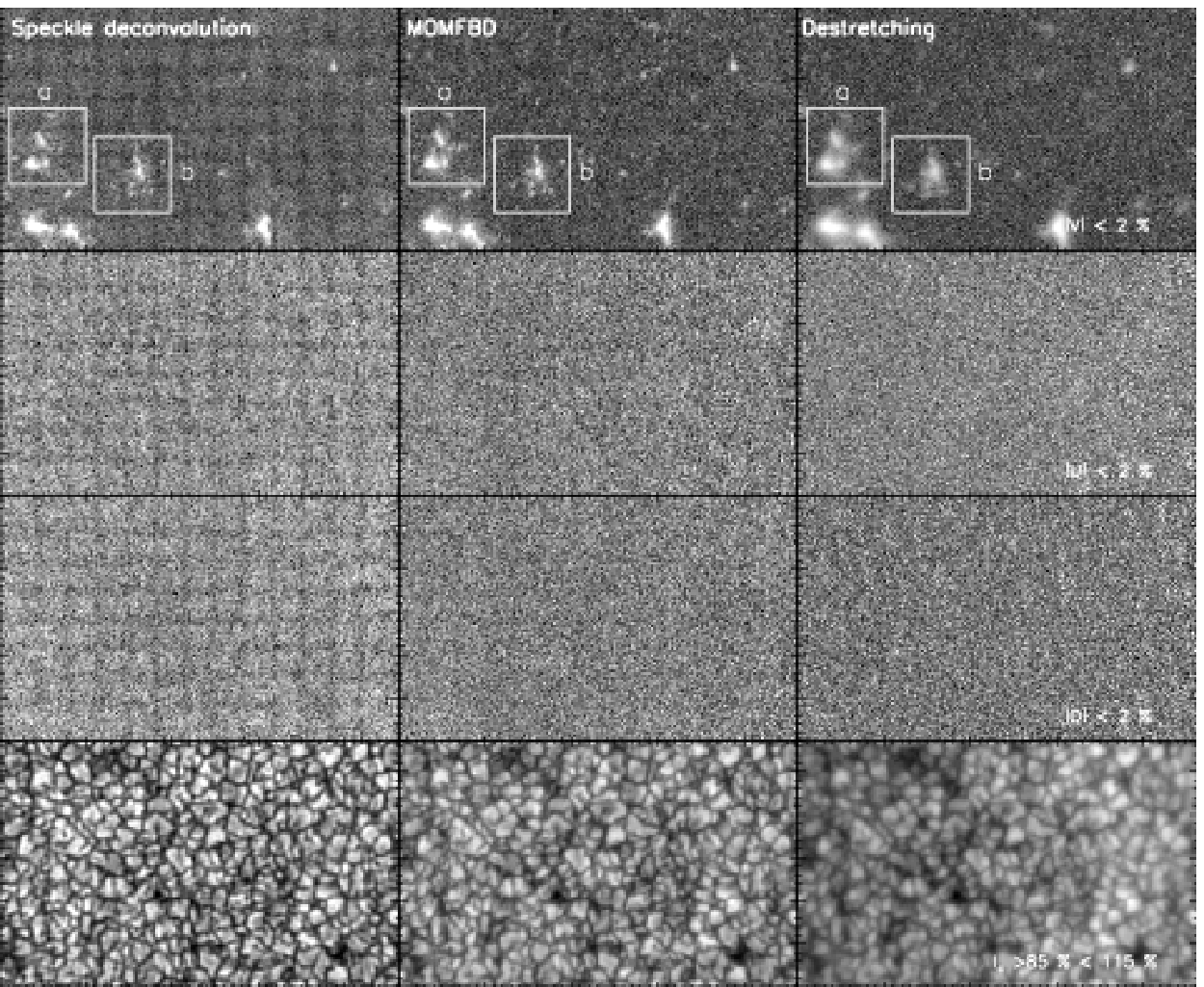}}}$ $\\
\caption{Overview maps of the NB data. {\em Left to right}: speckle deconvolution, MOMFBD, and destretching. {\em Bottom to top}: 0.85 $<I_c<$1.15, wavelength-averaged unsigned Stokes $Q < 2\%$, $U < 2\%$, and $V < 2\%$. The polarization maps are displayed on a logarithmic scale.  \label{fig_nb_overview}}
\end{figure*}

Figure \ref{fig_spec_filter} shows example spectra of Stokes $IQUV$ from  a location inside the network ({\em top panel}) and a small isolated polarization patch ({\em bottom panel}) for all three data reduction methods
before and after the application of the two filters. The low-pass filter and
the spectral running mean produce identical spectra. The largest noise
peaks in the speckle-deconvolved data are in some cases not fully removed by
the filtering, but spectrally broadened such that they resemble a false solar polarization signal with an  amplitude of up to 1\,\% of Stokes $I$ in some cases (e.g., in the {\em black rectangle} in $Q$ in the {\em uppermost row} of Fig. \ref{fig_spec_filter}). Some of the peaks, e.g., in Stokes $V$ of the speckle-deconvolved data in the {\em lower panel} of Fig. \ref{fig_spec_filter}, are located at wavelengths that would require Doppler shifts of several tens of kms$^{-1}$ or field strengths far exceeding 2 kG. These features are also missing in the other two reduction methods. This excludes a solar origin, but indicates that it is not straightforward to separate real from false polarization signals at the noise level of the speckle-deconvolved data (see Sect.~\ref{spectropol_prop} below).

The shape of the intensity profile stays virtually the same for all data reduction methods and applied filters. Table \ref{tab_noise} lists the rms noise in Stokes $V/I$ in a continuum window before and
after noise filtering. Because the weighted running mean and the low-pass filter have nearly the same effect on the spectra, we selected the weighted running mean as our method of choice for noise filtering. It should have the lowest impact on the spectral properties because it only operates on directly neighboring wavelength steps. 
\section{Results \label{sect_results}}
\subsection{Overview maps}
For a first visual inspection of the results, Figs.~\ref{fig_imaging_overview} and \ref{fig_nb_overview} show intensity maps of the reconstructed imaging data (blue continuum, \ion{Ca}{ii} H, 630\,nm BB data) and all Stokes parameters of the spectropolarimetric data, respectively.
\paragraph{Imaging data} In case of the imaging data in the blue continuum and
Ca ({\em upper two rows} in Fig.~\ref{fig_imaging_overview}), the two
reconstruction methods yield generally similar results with only a slightly
different behavior in the two wavelength channels. For this specific burst,
the MFBD of the Ca data has a slightly better spatial resolution as can be
seen when comparing the shape of the brightenings around $(x,y)\approx
(15^{\prime\prime},8^{\prime\prime})$ in the speckle and MFBD data that is
displayed magnified on the {\em right-hand side} of
Fig.~\ref{fig_imaging_overview}. The MFBD resolves three individual round
brightenings inside the {\em red square} in the magnification, whereas the
speckle reconstruction shows the ``blooming'' effect, i.e., a radial smearing
of isolated bright features. This effect produces a bright ring around a
central darkening in the difference image. Some more examples throughout the
FOV are marked by the {\em small grey squares} in the Ca difference image. The
burst shown was randomly picked near the end of the time-series, but a
comparison of the other reconstructed Ca images showed always the same trend,
i.e., a slightly better performance of MFBD in Ca in terms of spatial
resolution. For the blue continuum data ({\em middle row}), the reconstruction
results of both methods are virtually identical. The structures in the
difference image are usually located at the edges of granules and come from
the slightly different placing of individual subfields on a subpixel
scale. For the reconstruction of the BB data at 630\,nm taken with the
G{\"o}ttingen FPI ({\em bottom row} of Fig.~\ref{fig_imaging_overview}), the
results differ unexpectedly strong. The speckle reconstruction shows a
significantly larger intensity contrast than the MOMFBD. The contrast
difference reproduces the complete granulation pattern in the difference image (see Fig.~\ref{test_filter} for the trade-off between noise level and contrast in the MOMFBD).
\paragraph{Spectropolarimetric data} For the spectropolarimetric data of the
G{\"o}ttingen FPI, Fig.~\ref{fig_nb_overview} shows maps of the four Stokes parameters. The map of the continuum intensity $I_c$ was taken at the first wavelength step of the spectral scan. For Stokes $QUV$, the maps of the wavelength-averaged unsigned polarization signal, $\int |QUV/I| (\lambda) d\lambda / \int d\lambda$, are displayed. As for the BB data, the speckle
deconvolution has a significantly larger contrast than MOMFBD in $I_c$. No location inside the observed FOV shows a significant linear polarization signal ({\em second} and {\em third row} of Fig.~\ref{fig_nb_overview}). The noise level in the polarization states as given by the background patterns in $QUV$ is smallest for the destretched data set and largest for the speckle-deconvolved data. The latter also show a weak grid pattern at the noise level with an edge length of about 3$^{\prime\prime}$, which was also seen in \citet{sanchezandrade2009}. The reduced intensity of the grid pattern in the wavelength-averaged polarization signal implies a reduction of the noise level on those pixels where individual subfields were connected \citep{sanchezandrade2009}. The size of the grid matches the settings of the speckle reconstruction, i.e., subfields of 64 pixels ($\equiv 7.04^{\prime\prime}$) with a 33\,\% overlap to each neighboring one, where the apodization effects a gradual change from one subfield to the next. 

The polarization signals in Stokes $V$ show some differences between the different data reduction approaches. Several small-scale (roundish) features are well defined and resolved from each other in the speckle-deconvolved and MOMFBD data (e.g., inside the two {\em white rectangles}, especially in the lower part of rectangle {\em b}), whereas they appear significantly smeared out in the destretched data. The destretched data, however, show a polarization signal at the same locations as the deconvolved data sets. This implies that both deconvolution approaches in general do not reveal ``new'' signal that is absent in the destretched spectra, but only localize the origin of the polarization signal more precisely by removing the spatial smearing.
\begin{table}
\caption{Continuum contrast in percent.\label{tab_imaging}}
\begin{tabular}{ccccc}\hline\hline
Type & BB & NB & BC & CA \cr
 & 630\,nm & 630\,nm & 431\,nm & 396\,nm \cr\hline
Speckle & 8.0 & 6.3 & 7.6 & 6.4 \cr
(MO)MFBD & 4.6 & 4.3 & 7.8 & 6.5 \cr
Destretching &-- & 3.6 &-- & -- \cr
MHD &-- & 15.3  &-- & -- \cr
SP  &-- & 7.1   &-- & -- \cr\hline
\end{tabular}
\end{table}
\subsection{Continuum contrast and spatial power spectra\label{sect_spatres}}
We estimated the spatial resolution for the various maps quantitatively using the rms contrast and the power spectra of the intensity maps. The contrast gives only a partial estimate of the spatial resolution because it depends on various influences. Table \ref{tab_imaging} lists the rms contrast values. The numbers support the visual impression of the previous section: for the imaging channels, MFBD and speckle reconstruction give nearly identical contrasts. For the continuum intensity of the spectra or the BB channel, the contrast of the speckle deconvolution or reconstruction is larger by 2\,--\,4\% in absolute numbers (see Appendix \ref{appa} for other MOMFBD solutions). The observations have not been corrected for stray light yet, which will increase the absolute contrast values but presumably not their differences. Table \ref{tab_imaging} also lists the corresponding continuum intensity contrast in the MHD simulation and the Hinode/SP data (see Sect.~\ref{add_data}) for comparison.
\begin{figure}
\centerline{\resizebox{8.8cm}{!}{\includegraphics{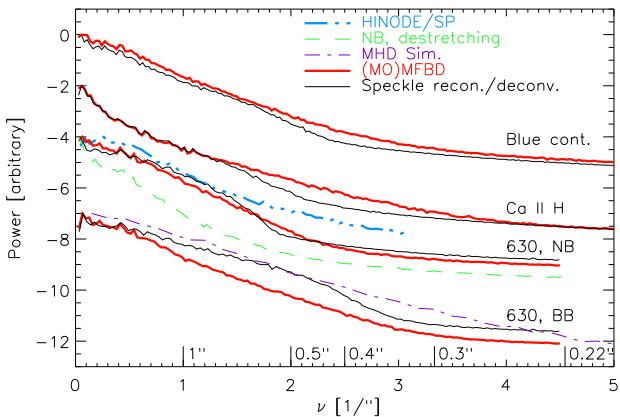}}}
\caption{Spatial power spectra. {\em Thick red}: (MO)MFBD data. {\em Thin black}: speckle-reconstructed/deconvolved data. {\em Top to bottom}: blue continuum, \ion{Ca}{ii} H imaging data, continuum window in the 630\,nm spectra, 630\,nm BB data (each vertically displaced by two units). The {\em thick blue triple-dot dashed}, {\em dash-dotted purple}, and {\em dashed green lines} show the power spectra of Hinode/SP 630\,nm spectra, a MHD simulation, and the destretched spectra, respectively. \label{spat_power_momfbd}}
\end{figure}

The spatial power spectra are shown in Fig.~\ref{spat_power_momfbd}. They are normalized individually to the power at the first non-zero frequency. The imaging data in blue continuum and Ca show a similar behavior, i.e., the power of the speckle reconstruction drops faster than the MFBD at spatial scales of about 0\farcs5 and levels off to a constant level earlier \citep[see also][]{matson2008}. This trend is more pronounced for the Ca data, as expected from the visual impression (Fig.~\ref{fig_imaging_overview}). For the spectra of the NB channel, the difference between the speckle deconvolution and the MOMFBD is small. The speckle deconvolution shows slightly larger power at spatial scales of 1$^{\prime\prime}$ and for scales smaller than 0\farcs4, where the latter produces the higher noise level of the speckle deconvolution. The power spectrum of the destretched spectra ({\em dashed green line}) lies below that of both deconvolved spectra, implying both a lower noise level and a worse spatial resolution. Both the MHD simulation ({\em red dash-dotted}) and the data from the Hinode/SP ({\em thick blue}) show a power spectrum with a nearly linear shape in the logarithmic plot that the power of the speckle-reconstructed BB images matches for spatial scales larger than about 0\farcs5. 

The effective spatial resolution is usually determined from the location where the power curve changes its slope significantly and levels off to a roughly constant level. The exact location, however, is difficult to pinpoint in observed power spectra. We would attribute a spatial resolution of about 0\farcs25 ($\equiv\nu=4$) to the imaging data and about 0\farcs4 to the deconvolved NB spectra, whereas the destretched NB spectra reach rather only about 0\farcs6. The power spectrum of the Hinode/SP data shows no real change of the slope, implying a diffraction-limited performance with 0\farcs32 as the theoretical resolution limit.
\subsection{Spectroscopic properties \label{spectropol_prop}}
To investigate a possible effect of the deconvolution methods on the shape of the spectra, we determined the following parameters: the average profile, the line width, the relative line asymmetry, the line-core intensity, and the line-core position (converted to the corresponding LOS velocity). 

\begin{figure}
\centerline{\resizebox{8.8cm}{!}{\includegraphics{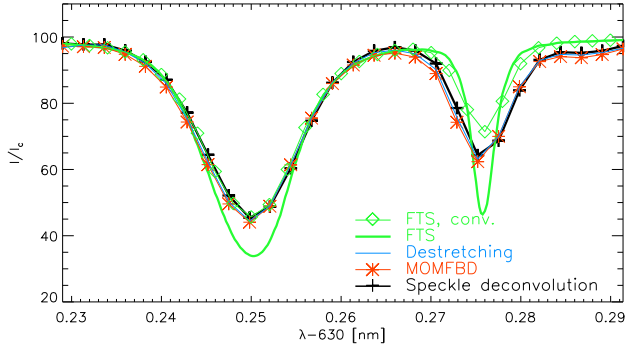}}}
\caption{Average profiles of the G{\"o}ttingen FPI. {\em Black line with crosses}: speckle deconvolution. {\em Red line with asterisks}: MOMFBD. {\em Blue line}: destretched data. The {\em thick  green} ({\em thin green line with diamonds}) line shows the corresponding FTS atlas profile before (after) the convolution with the instrumental profile.} \label{exam_prof}
\end{figure}

Averaging over the FOV yields nearly identical profiles for all data reduction
methods (Fig.~\ref{exam_prof}). Because the deconvolution should preserve the total intensity, and therefore also the average profiles, the very small differences imply that all three data reduction methods performed correctly in this respect. We matched the observed spectra and the corresponding section of the FTS atlas \citep{kurucz+etal1984} with the approach used in \citet{allendeprietoe+etal2004} and
\citet{cabrerasolana+etal2007} using a stray light offset $\alpha$ and a
subsequent convolution of the FTS spectrum with a Gaussian of width $\sigma$.
The Gaussian curve is assumed to account for the spectral resolution of
the instrument. The best-fit was reached for $(\alpha, \sigma)$ = (14\,\%,
1.65\,pm), which yields a FWHM of the Gaussian of 3.88\,pm and thus a spectral
resolution of about 162.000, somewhat below the theoretical limit of the G{\"o}ttingen FPI of about 240.000 \citep{nazi+kneer2008a}.
\begin{figure}
\centerline{\resizebox{8.5cm}{!}{\hspace*{0.3cm}\includegraphics{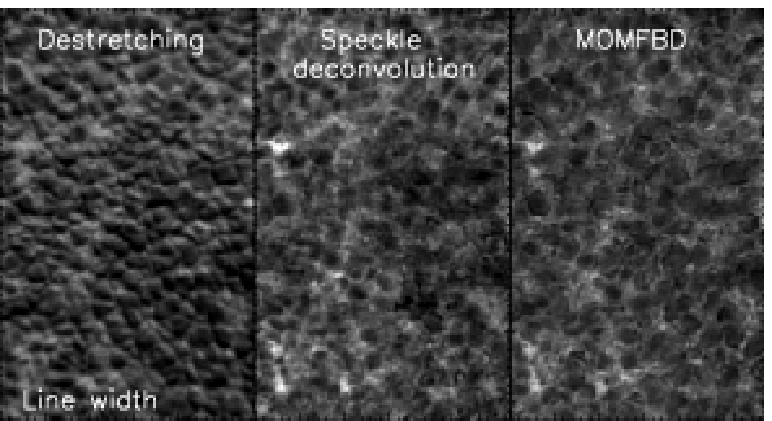}}}
\centerline{\resizebox{8.5cm}{!}{\hspace*{0.3cm}\includegraphics{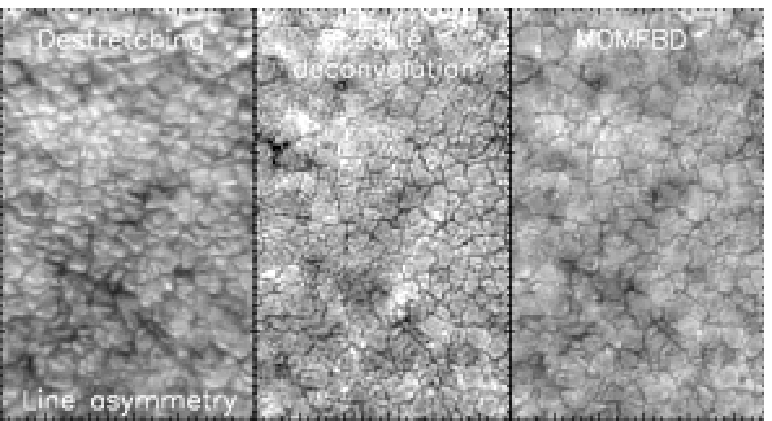}}}
\centerline{\resizebox{8.5cm}{!}{\hspace*{0.3cm}\includegraphics{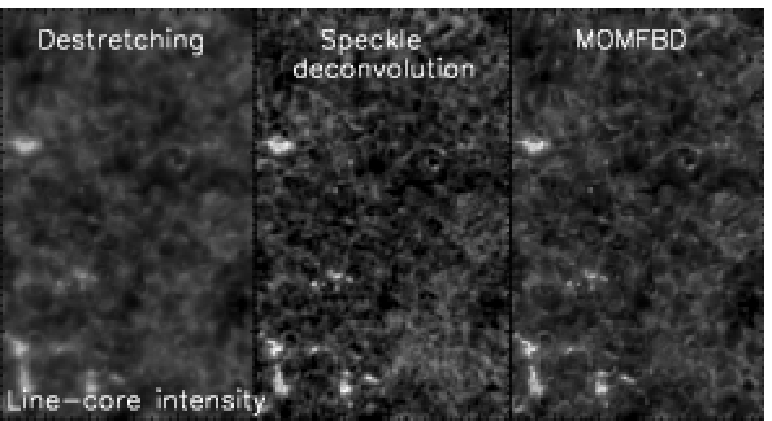}}}
\centerline{\resizebox{8.5cm}{!}{\hspace*{0.3cm}\includegraphics{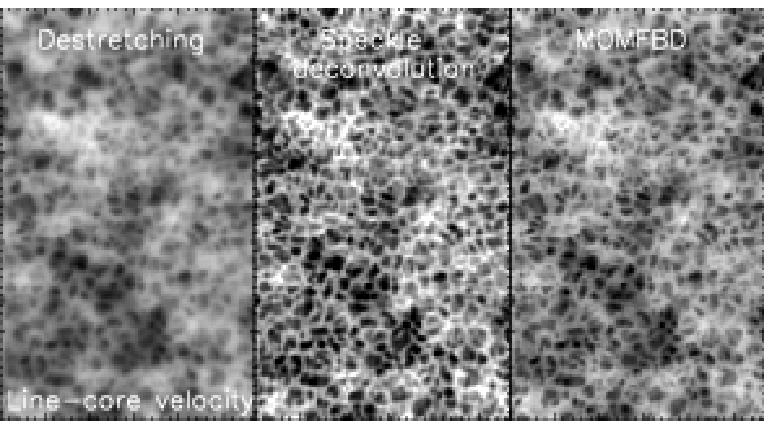}}}$ $\\
\centerline{\resizebox{8.5cm}{!}{\hspace*{0.3cm}\includegraphics{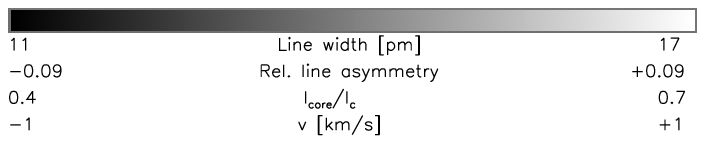}}}$ $\\

\caption{Line parameters. {\em Top to bottom}: line width, line asymmetry,
  line-core intensity, and line-core velocity. {\em Left to right}:
  destretched data, speckle-deconvolved data, MOMFBD data. Tick
  marks are in arcsec. The display ranges are given by the gray-scale bar at bottom.\label{lineparam_2d}}
\end{figure}

Prior to the determination of the other line parameters, we removed the
telluric line blend at 630.27\,nm from the spectra by fitting a Gaussian to it
and subtracting the corresponding contribution from the profile. The line
width was provided by a subsequent Gaussian fit to the solar line. The
line-core position of the solar line was derived using a Fourier method for
the close surroundings of the intensity minimum. The position was finally
converted to the corresponding LOS velocity using the average line-core
position as zero reference. The line asymmetry was determined as the
difference of the area between the continuum intensity level and the line profile to the red and the blue from the core position for an equal distance of $\pm 16.2$\,pm in wavelength. The difference was then normalized with the total area. Prior to the
  determination of the line asymmetry, the whole profile was shifted to have the line-core position exactly at one of the observed wavelength steps to avoid
spurious asymmetry values caused by the discrete wavelength sampling.  The line-core intensity was derived from the minimum intensity in the spectral line.
\begin{table}
\caption{Line parameters of the intensity profiles and their rms values. \label{tab_lineparam}}
\begin{tabular}{ccccc}\hline\hline
Data & line width & line asym.& core int. & $v_{\rm LOS}$ \,rms \cr
    &  pm &  \% & \% of $I_c$ & km\,s$^{-1}$ \cr\hline
Destretching & 12.33$\pm$0.71 &1.5$\pm$2.4 & 46.1$\pm$2.2 & 0.296 \cr
MOMFBD    & 12.47$\pm$0.57 &2.1$\pm$1.9 & 45.7$\pm$2.6 & 0.347 \cr
Speckle deconv.   & 12.57$\pm$0.69 &2.7$\pm$2.6 & 45.5$\pm$3.2 & 0.454 \cr
Hinode/SP & 11.71$\pm$0.99 &$-0.2\pm$1.4& 38.8$\pm$5.1 & 0.537 \cr
MHD Sim.  & 12.18$\pm$1.21 &$-0.1\pm$2.5& 38.3$\pm$7.6 & 1.029 \cr
MHD, degr. & 12.43$\pm$0.76 &$-0.6\pm$1.2& 40.2$\pm$5.0 & 0.728 \cr\hline
\end{tabular}
\end{table}

Figure \ref{lineparam_2d} shows maps of the retrieved line
parameters. The image orientation is at 90 degrees to the previous
  figures. The line-core intensity and line-core velocity show a similar pattern for both deconvolution techniques and a considerably improved resolution when compared with the destretched data set. Locations of strong magnetic fields can be clearly identified in the line-core intensity \citep[see
also][]{puschmannetal07} and the line-width map. For both the line asymmetry and the line width, the deconvolution changed, however, to some extent the spatial structuring across the FOV when compared with the destretched data set. A weak grainy pattern can be identified in the line width of the speckle-deconvolved data that was noted as typical for this reconstruction method also by \citet{mikurda+etal2006}.

Table \ref{tab_lineparam} lists the average values of the line parameters and their rms fluctuations for the G{\"o}ttingen FPI spectra, the Hinode/SP data set, and the MHD simulation. The table is roughly sorted by increasing spatial resolution from top to bottom (cf.~Sect.~\ref{sect_spatres}), except for the last line that corresponds to the MHD simulation after a spatial degradation with a 0\farcs52 wide Gaussian kernel, which intends to mimic the resolution of the POLIS data. For most quantities, e.g., LOS velocity, line width, and line-core intensity, a clear trend of increasing rms variations can be seen with the spatial resolution. The larger line width and higher line-core intensity of the G{\"o}ttingen FPI spectra in comparison to the other two data sets are presumably caused by the instrumental profile and the resulting lower spectral resolution of the G{\"o}ttingen FPI. 
\begin{figure}$ $\\$ $\\
\centerline{\resizebox{8.8cm}{!}{\hspace*{.4cm}\includegraphics{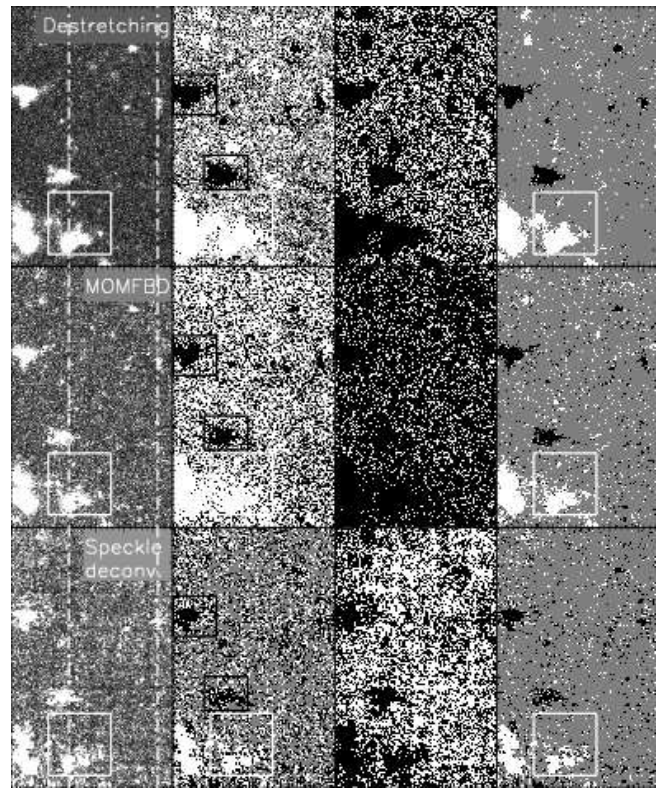}}}$ $\\
\caption{{\em Left to right}: maximal $V$ amplitude thresholded at $1.5\%$, polarity, locations with $V_{\rm max} > 2.5\,\sigma$ ({\em black}), and polarity of the significant polarization signals with the final rejection thresholds. {\em Top to bottom}: destretched data, MOMFBD, speckle-deconvolved data. The {\em white rectangle} marks the area that is displayed magnified in Fig.~\ref{mag_pol_prop}. The two {\em vertical dash-dotted lines} denote the location of the two cuts shown in Fig.~\ref{spec_cuts}. \label{spec_prop1}}
\end{figure}
\subsection{Spectropolarimetric properties}
The intensity spectra showed no strong effect by the deconvolution, but the spectra still have a considerable noise level even after the spectral filtering. We thus first focus on quantities that allow to determine the significance level of the polarization measurements like the Stokes $V$ amplitude and the polarity of the magnetic fields before we turn to the net circular polarization (NCP). 

We used the maximum value of the unsigned polarization signal in Stokes $V$ as
proxy for the presence of a polarization signal because the solar Stokes $Q$ and $U$ signals are below the noise. The polarity was defined as the order of
maximum and minimum signal in the signed Stokes $V$ profile and set to $\pm 1$
accordingly. We used a threshold of 2.5 times the rms noise level $\sigma$ in
the Stokes $V$ spectra (see the last column of Table\ref{tab_noise}) as
initial criterion for a significant signal \citep[cf.~the 2\,$\sigma$ level
used in parts of][]{nazi+kneer2008a}. The polarity was set to zero for all
locations with $V_{\rm max} < 2.5\,\sigma$. Figure \ref{spec_prop1} shows the
maps of the Stokes $V$ amplitude, the polarity, and the mask of locations with $V_{\rm max} > 2.5\,\sigma$ ({\em black shading}).
\begin{figure}$ $\\$ $\\
\centerline{\resizebox{8.8cm}{!}{\hspace*{.5cm}\includegraphics{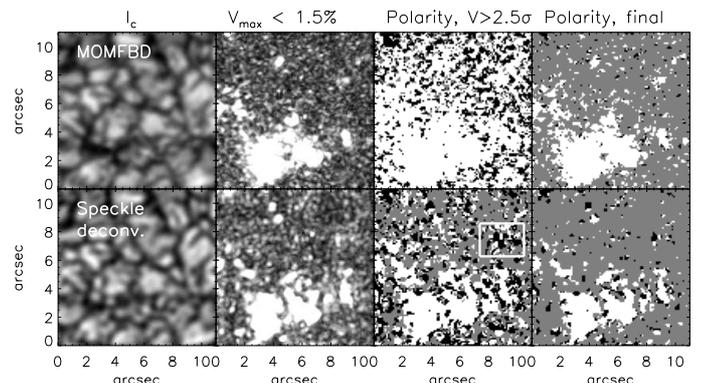}}}$ $\\
\caption{Magnification of the area marked in Fig.~\ref{spec_prop1}. {\em Left
    to right}: continuum intensity, maximal $V$ amplitude $< 1.5\%$, polarity
  for $V> 2.5\,\sigma$, and polarity with the final rejection thresholds. {\em Top}: MOMFBD. {\em Bottom}: speckle deconvolution. \label{mag_pol_prop}}
\end{figure}

From a comparison of the masks or the polarity maps with the Stokes $V$ amplitude, it is clear at a first glance that the 2.5\,$\sigma$ threshold  is insufficient because in many places no significant corresponding signal can be seen in the Stokes $V$ map. The polarity of the destretched and MOMFBD data matches for most locations where the signal level in the $V$-amplitude map exceeds the display threshold of 1.5\,\% (e.g., inside the {\em black rectangles}). The speckle-deconvolved data sometimes show a polarity opposite to the other two data sets (e.g., at some places inside the {\em lower black rectangle}). The polarities in both deconvolved data sets tend to fluctuate on small spatial scales at the borders of large-scale connected polarization patches \citep[cf.~also][their Sect.~9.3]{pillet+etal2011}.

These fluctuations of the polarity continue in the areas with weak or no polarization signals, but the three data reduction methods show a slightly different behavior there. The destretched and MOMFBD spectra preferentially show a salt'n'pepper pattern with pixel-to-pixel variations typically for random noise (Fig.~\ref{spec_prop1} and the magnification in Fig.~\ref{mag_pol_prop}). The speckle-deconvolved spectra on the other hand in some cases exhibit a very regular pattern of repeatedly alternating polarities extending over patches of 2\,--\,3 pixels each, for instance inside the {\em white rectangle} in the polarity of the speckle-deconvolved spectra in Fig.~\ref{mag_pol_prop}. This regular pattern resembles a few rows or columns from a ``chequerboard'' \citep{nazi+etal2009}, but its regularity is rather surprising in the dynamic QS environment. Because the pattern is less prominent in the MOMFBD data and almost completely absent in the destretched data, we tried to trace down its origin using the spectra along the spatial cuts marked by the {\em vertical white dash-dotted} lines in Fig.~\ref{spec_prop1}. 
\begin{figure*}
\begin{minipage}{11cm}$ $\\
\centerline{\resizebox{11.cm}{!}{\hspace*{.4cm}\resizebox{2.75cm}{!}{\includegraphics{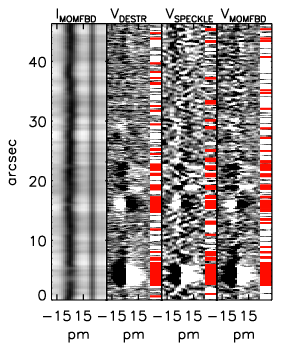}}\hspace*{.75cm}\resizebox{2.75cm}{!}{\includegraphics{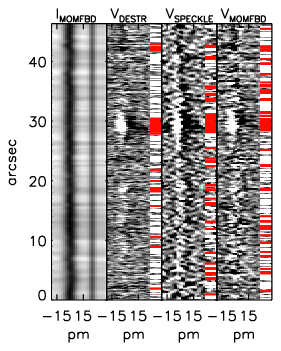}}\hspace*{.75cm}\resizebox{2.06cm}{!}{\includegraphics{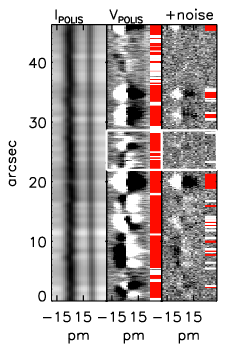}}}}$ $\\
\end{minipage}\hspace*{.6cm}
\begin{minipage}{6.cm}
\caption{Stokes $I$ and $V$ spectra along the two cuts marked in
  Fig.~\ref{spec_prop1} ({\em left two panels}), and in a reference data set taken with POLIS ({\em
    right panel}). {\em Left panel, left to right}: intensity spectrum, Stokes $V$ of the destretched, speckle-deconvolved, and MOMFBD data, respectively. The display range of $V$ is $\pm 0.5\%$. {\em Middle panel}: the same for the second cut. {\em Right panel, left to right}: $I$ and $V$ ($\pm 0.1\%$) from the POLIS data, $V$ plus 0.2\% rms noise with a display threshold of $\pm 0.5\%$. The {\em white markers} at the {\em right border} of Stokes $V$ denote $V_{\rm max} > 2.5\sigma$, the {\em red markers} $V_{\rm max}$ above the finally chosen significance level.\label{spec_cuts}}
\end{minipage}
\end{figure*} 

The Stokes $I$ and $V$ spectra along these cuts are displayed in
Fig.~\ref{spec_cuts}. Comparing the $V$ profiles of the destretched data ({\em
  second column} in the {\em left two panels}) with those of both
deconvolved data sets ({\em third} and {\em fourth column}), one can detect again a similar black-and-white chequerboard pattern in the deconvolved
spectra, but this time the $x$-axis corresponds to the {\em wavelength}, not
a spatial dimension. Most of these features that reach and exceed the display
threshold of $\pm 0.5\,\%$ amplitude are seen on 2\,--\,3 wavelength points, whereas the line width is larger than five wavelength steps because of the thermal
and Doppler broadening, and over 2\,--\,3 pixels in the spatial dimension. The pattern therefore is generated by the reconstruction of the individual wavelength points and is not caused by any solar polarization signal.

Its artificial origin is demonstrated also by the comparison to the {\em right panel} of Fig.~\ref{spec_cuts}, which shows a cut of identical length through the low-noise POLIS data set (see Sect.~\ref{add_data}). The {\em second column} of the {\em rightmost panel} of Fig.~\ref{spec_cuts} shows the original Stokes $V$ spectrum with a threshold of $\pm 0.1\%$. The Stokes $V$ spectrum shows a polarization signal all along the slit, where all
weak signals still have a larger spectral extent than the black/white pattern in the deconvolved G{\"o}ttingen FPI spectra. The {\em rightmost column} of Fig.~\ref{spec_cuts} displays the same spectrum after an addition of a rms noise of 0.2\%, which is
typical for the G{\"o}ttingen FPI data, with the same display range of $\pm 0.5\%$ as for the G{\"o}ttingen FPI data. The added noise pushes about 40\,\% of the previously significant signals below the noise level, as for instance those inside the {\em white rectangle} near the middle of the cut. The clearly different appearance of the noise in the deconvolved spectra and the rightmost Stokes $V$ spectrum suggests that the noise in the deconvolved spectra should not be treated as being Gaussian. It shows a spectral and spatial coherence over a few pixels in both dimensions caused by the noise filtering and the reconstruction process, respectively. The noise in the destretched data deviates from a Gaussian noise as well because of the spectral filtering, but rarely exceeds
the display threshold of 0.5\,\% with noise peaks.
\begin{table}
\caption{Final significance threshold in percent ({\em top row}) and relative to the rms noise $\sigma$ in Stokes $V$ ({\em middle row}). Area coverage of polarization signals above the threshold ({\em bottom row}). \label{tab_sign}}
\begin{tabular}{cccccc}\hline\hline
 & Speckle & MOMFBD &  Destret- & POLIS &POLIS\cr
 & deconv. &  & ching  &  &+noise \cr\hline
\%& 1.1 & 1.0 & 0.8 & 0.11  & 0.6 \cr
$\sigma\;\times$ & 3.5 & 5 & 4.5 & 3.5 & 3 \cr
\% of area & 18 & 16 & 13 & 73 & 35 (13\tablefootmark{1})\cr\hline
\end{tabular}
\tablefoottext{1}{after extrapolation to the spatial sampling of the G\"ottingen FPI}
\end{table}

We then determined suitable significance thresholds for the various data sets using the Stokes $V$ spectra of Fig.~\ref{spec_cuts} as a guide. Profiles that exceeded the initially chosen 2.5\,$\sigma$ level are indicated by the {\em white markers} at the right border of all Stokes $V$ spectra. This clearly selects many locations without any solar polarization signal in the spectra. The {\em red markers} correspond to locations that remain with the manually selected thresholds listed in Table \ref{tab_sign}. These signals can mostly be confirmed to be significant by eye as well. 

\begin{figure}
\centerline{\hspace*{.9cm} POLIS \hspace{.8cm} Hinode/SP \hspace{.8cm} MHD Sim.\hspace*{1.5cm}}$ $\\
\centerline{\hspace*{.7cm}\resizebox{1.8cm}{!}{\includegraphics{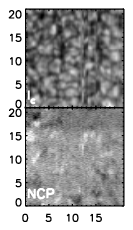}}\hspace*{.4cm}\resizebox{1.8cm}{!}{\includegraphics{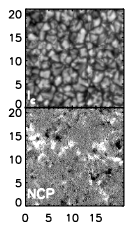}} \hspace*{.3cm}\resizebox{1.8cm}{!}{\includegraphics{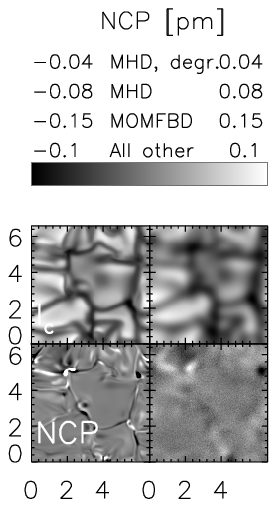}}\hspace*{1.8cm}}$ $\\$ $\\
\centerline{\hspace*{.7cm}\resizebox{8.cm}{!}{\includegraphics{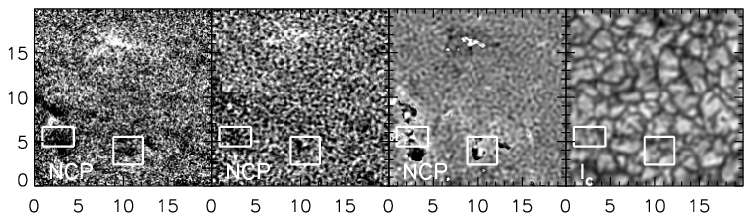}}}$ $\\$ $\\
\centerline{\hspace*{.6cm} Destretching \hspace{.3cm} MOMFBD \hspace{.3cm} Speckle deconv.\hspace{2.cm}}
\caption{NCP and intensity maps of the data sets. {\em Bottom row, left to
    right}: maps of the NCP for the destretched, MOMFBD, and
  speckle-deconvolved G{\"o}ttingen FPI spectra, and continuum intensity. {\em Top left}: continuum intensity ({\em top panel}) and NCP for the POLIS data set. {\em Top middle}: same for the Hinode/SP data set. {\em Top right}: continuum intensity and NCP for the MHD simulation in full ({\em left}) and degraded resolution ({\em right}). All tick marks are in arcsec.\label{ncp_fig}} 
\end{figure}

The {\em rightmost columns} of Figs.~\ref{spec_prop1} and \ref{mag_pol_prop} show the polarity of the locations of significant polarization signals inside the FOV of the G{\"o}ttingen FPI when the rejection thresholds are applied. They are now mainly limited to the signals that can be clearly identified in the map of the Stokes $V$ amplitude ({\em leftmost column}). The chequerboard pattern has disappeared almost completely from both deconvolved data sets. The fraction of the FOV that shows significant polarization signal is about 15\,\% for the G{\"o}ttingen FPI data, in comparison to about 70 (30)\% for the POLIS data (+noise) ({\em bottom row} of Table \ref{tab_sign}).
\paragraph{Stokes $V$ amplitudes}
The reconstruction improves the spatial resolution and also amplifies both the polarization amplitudes and the noise level. Scatter plots of only the significant polarization signals showed an increase of the polarization amplitude relative to the destretched data by 20\,\% for the MOMFBD and 40\,\% for the speckle deconvolution (cf.~also Fig.~\ref{v_ampl_comp}). The increase of the polarization amplitude implies that the magnetic structures are better resolved in the deconvolved data sets with a higher filling factor inside a pixel. The amount of increase is in rough agreement with the ratio of the spatial resolution estimated in Sect.~\ref{sect_spatres} (0\farcs4/0\farcs6 = 2/3, i.e., a 33\,\% smaller effective area per pixel for the deconvolved spectra).
\paragraph{Net circular polarization} The net circular polarization (NCP) is
an important diagnostic quantity because it indicates gradients along the LOS in
the magnetic field and the velocity field \citep[see, e.g.,][]{almeida+lites1992}. We defined the NCP as $\int V d\lambda$ ignoring
the polarity. Figure \ref{ncp_fig} compares the NCP maps of the various data
sets: the three reductions of the G{\"o}ttingen FPI spectra ({\em bottom row}),
the POLIS and Hinode/SP observations, and the MHD simulation in full
resolution and spatially degraded to 0\farcs52 resolution ({\em top row} of
Fig. \ref{ncp_fig}). The general\footnote{The
  observations are neither simultaneous nor co-spatial.} shape of structures in the NCP of the  slit-spectrograph observations (POLIS, Hinode/SP) matches that in the MHD simulation with a spatial pattern on granular scales. For the destretched G{\"o}ttingen FPI spectra, the NCP maps are rather noisy, but compare well to the Hinode/SP data set when noise of the level of the G{\"o}ttingen FPI spectra would be added to the latter, showing connected patches of non-zero NCP values. 
\begin{figure}
\centerline{\resizebox{8.8cm}{!}{\includegraphics{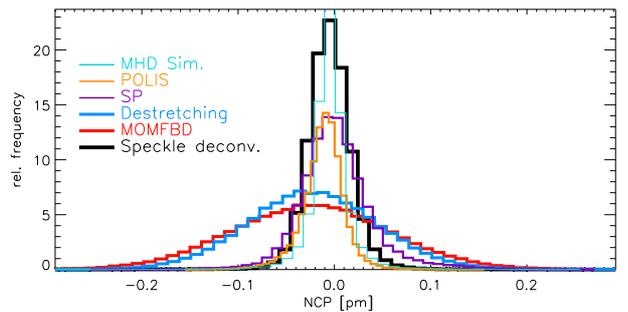}}}
\caption{Histograms of the NCP values in the destretched data ({\em thick
    blue}),  MOMFBD data ({\em thick red}), and  speckle-deconvolved data
  ({\em thick black}). For comparison, histograms for Hinode/SP spectra ({\em thin purple}), POLIS spectra ({\em thin orange}), and for spectra from a MHD simulation ({\em thin turquoise}) are also shown.\label{ncp_fig1}}
\end{figure}

The NCP map of the destretched spectra shows pixel-to-pixel variations outside the large-scale connected polarization patches, whereas both the MOMFBD and speckle-deconvolved spectra show a distinct background pattern of individual circles of a few pixel extent, more clearly seen for the speckle-deconvolved spectra. The noise level in the spectra affects the NCP strongly, such thus for instance in the {\em left white rectangle} of Fig. \ref{ncp_fig} the values from the three data reduction methods contradict each other: the destretched data suggest a negative NCP (black), in the MOMFBD spectra the signal is basically noise, and the speckle-deconvolved spectra exhibit a positive NCP (white) there. The general noise level in the NCP of the speckle-deconvolved spectra is surprisingly smaller by an order of magnitude compared to the other two data reduction methods, even if the noise in the NCP should scale with the noise level in the spectra. The histograms of the NCP values in Fig.~\ref{ncp_fig1} show that the distribution of the NCP in the speckle-deconvolved spectra ({\em thick black}) matches better to all slit-spectrograph observations and the MHD simulation, whereas the destretched and MOMFBD spectra ({\em blue} and {\em red} curves) have much broader distributions. On closer inspection, the borders of the NCP patches in the speckle-deconvolved data seem to be artificially sharp (e.g., in the {\em right white rectangle}), changing from about zero to the display threshold from one pixel to the next. The same behavior of the NCP also resulted for those MOMFBD test runs where the NB channel was not fully weighted in the deconvolution (see Fig.~\ref{ncp_comp_test}).

\begin{figure}
\centerline{\resizebox{8.8cm}{!}{\includegraphics{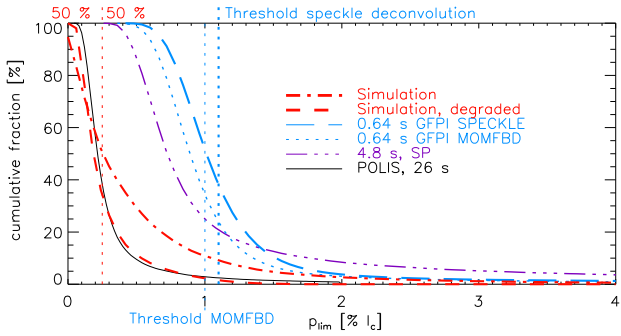}}}
\caption{Cumulative fraction of profiles with a polarization degree
  above $p_{\rm lim}$. {\em Thin black}: POLIS data with
  26 secs integration time. {\em Purple thin dash-dotted}: Hinode/SP. {\em Blue dotted}: MOMFBD G{\"o}ttingen FPI data. {\em Blue long dashed}: speckle-deconvolved G{\"o}ttingen FPI data. {\em Thick red dash-dotted}: MHD simulation. {\em Thick red dashed}: spatially degraded MHD simulation. The {\em vertical red dotted line} denotes the 50\% fraction of the MHD simulation. The {\em vertical blue dotted lines} mark the significance threshold for the MOMFBD and speckle-deconvolved spectra.\label{cumulative_frac}}
\end{figure}

The reason for this behavior of the NCP in some of the deconvolutions is not clear at once. The NCP is determined from the addition of several wavelength points, making it strongly sensitive to individual noise peaks. The shape of the distribution (Fig.~\ref{ncp_fig1}), however, suggests that the noise in the NCP of, e.g., the speckle-deconvolved data is closer to that of observations with higher S/N ratio than that of the other two data reduction methods, even if the rms variation in the spectral dimension is higher for the speckle-deconvolved data. Adding up only the polarization signal in the blue or red wing of the line yielded nearly identical images for all data reduction methods, the sharp boundaries only appeared as soon as the two images were subtracted. We investigated the spatial Fourier power of the NCP maps, but this unfortunately also did not provide a clear clue to the observed behavior of the NCP. 
\paragraph{Cumulative polarization fraction}
Another method to compare data from different sources is the cumulative
fraction of all profiles that exceed a given polarization limit $p_{\rm
  lim}$. This fraction is a rather ``robust'' quantity because it is based on good statistics, and at the same time allows directly to see the detection
limit set by the noise level \citep[see also Fig.~5 in][]{beck+rezaei2009}. The noise level in the polarization degree can be
read off from the location, where the curves start to deviate from
100\%. Figure~\ref{cumulative_frac} shows the result for the G{\"o}ttingen FPI, POLIS, and Hinode/SP observations, as well as the original MHD simulation and the MHD simulation after degradation to a resolution of about 0\farcs52.

The respective curves and noise values generally follow the trend predicted by their integration time $t$, which is defined here as the total exposure time that entered into the measurement of the Stokes vector at one wavelength step (G{\"o}ttingen FPI) or one slit position (POLIS, Hinode/SP). In comparison with the original and degraded MHD simulation ({\em thick red curves} in Figure~\ref{cumulative_frac}), only the POLIS observation with $t\,=\,26$ secs is able to measure the bulk of the polarization signals in the MHD simulation. For the Hinode/SP
data ($t\,=\,4.8$ secs), only the stronger half of the polarization signals of the MHD simulation ({\em red vertical dotted line} in Figure~\ref{cumulative_frac} at the 50\,\% level) are just at or slightly below the noise level. For the speckle-deconvolved/MOMFBD G{\"o}ttingen FPI data ($t\,=\,0.64$ secs), the cumulative fraction reduces to values below 100\,\% for a $p_{lim}$ of about 0.5\,\% of $I_c$. Thus, only 30\,\% of the polarization signals present in the MHD simulations are above the noise level and can be detected. Only about 8\,\% (10\,\%) of the polarization signals in the MHD simulation are above the significance threshold needed for the speckle-deconvolved (MOMFBD) G{\"o}ttingen FPI spectra ({\em vertical blue dotted lines} in Figure~\ref{cumulative_frac}). The signals in the simulation, however, correspond to the comparably weak magnetic fields on granular scales. The G{\"o}ttingen FPI and the Hinode/SP data show  more locations with large polarization signals ($p_{\rm lim}> 1 \%$) than the POLIS data, which results from their higher spatial resolution: for unresolved magnetic fields inside a pixel, the polarization degree increases with spatial resolution because it measures the fraction of polarized and unpolarized photons.
\section{Summary and Discussion\label{sect_summ}}
We have applied two different image reconstruction techniques (speckle
reconstruction/deconvolution, (MO)MFBD) to  imaging and 2D spectropolarimetric data. For the imaging data (blue continuum at 431.3\,nm and \ion{Ca}{II} H), both reconstruction methods yielded similar results with nearly identical rms contrasts, with a slightly better performance of the MFBD approach in the case of \ion{Ca}{II} H imaging data with a low light level. For the reconstruction of both the BB data (at 630\,nm) or the NB data (\ion{Fe}{i} line at 630.25\,nm) taken with the G{\"o}ttingen FPI, the results differed significantly. The speckle approaches yielded a significantly higher continuum rms contrast (6.3\,\%) than the MOMFBD with equal weight for BB and NB channel (4.3\,\%), with a similar spatial resolution for both methods. 

A series of test deconvolutions (Appendix \ref{appa}) showed a clear trade-off between contrast and noise level in Stokes $V$, where the solution of a MOMFBD with equal weights for BB and NB channel turned out to be the one with the lowest noise, whereas the speckle-deconvolution corresponds to one with medium noise and medium contrast. For a derivation of the magnetic properties of the solar photosphere, the solution with lowest noise is to be preferred because the intensity contrast will enter stronger into the spatial variation of the derived temperatures than into the magnetic field properties. One would, however, expect more similar results for the speckle deconvolution and the MOMFBD because the methods were applied to the same data set, although the final outcome depends strongly on the settings during the deconvolution (Appendix \ref{appa}). One reason for the difference between the speckle deconvolution and the MOMFBD intrinsic to the data used could be (small) differences in the optical path between BB and NB channel in the G\"ottingen FPI that the MOMFBD reconstruction is more sensitive to than the speckle technique (M.~L{\"o}fdahl, personal note). Another reason could be uncompensated high-order aberrations \citep{rouppe+etal2004,scharmer+etal2010} that influence the contrast more than the spatial resolution.

Comparing the deconvolved intensity spectra with a data set that was
only corrected by applying a destretching algorithm, both deconvolution
methods increased the spatial resolution without introducing prominent data
artifacts in the intensity profiles. ``Artifacts'' in this context would be all patterns whose properties are not compatible with a solar origin, such as pixel-to-pixel variations above the limit given by the noise level (because at the spatial resolution of the data solar features should cover consistently a few pixels), and small or large-scale patterns in the deconvolved spectra other than those seen also in the destretched spectra (e.g., spatially extended solar oscillations have to show up also without the deconvolution). Such cases are not seen in the line parameters shown in Fig.~\ref{lineparam_2d}. The two parameters with a pronounced difference between destretched and deconvolved spectra are the line width and, to a lesser extent, the line asymmetry. The reason why these quantities change more than, for instance, the line-core intensity, presumably is that both are non-linear measures, where one additionally also expects a large intrinsic difference of the values inside granules or intergranular lanes. The increase of the spatial resolution by the deconvolution therefore has a profound effect on these two quantities, but the change seems to be of solar origin. The grainy pattern in, e.g., the line width of the speckle-deconvolved data is so weak that it does not qualify as significant because its amplitude will presumably not be noticeable in an analysis of the data.

An investigation of the polarization signal revealed, however, one of the drawbacks of the deconvolution, i.e., the amplification of the noise level. All of the spectra have to be noise filtered in the spectral domain to be useful. For the speckle-deconvolved data, spurious polarization signals of up to 0.5-1.0\,\% of $I_c$ remained even after a rather strong filtering. The high noise level in the deconvolved data can affect the analysis of the polarization signal producing for example a chequerboard pattern in the polarity of the magnetic fields that is absent in the destretched data. The pattern could be traced back to individual noise peaks in the Stokes $V$ spectra that are not caused by a solar polarization signal. From a visual inspection of the spectra and a comparison to a reference slit-spectrograph spectrum with a noise level lower by one order of magnitude, we determined significance thresholds for the deconvolved data.

For the speckle deconvolution (MOMFBD), signals with a polarization amplitude below about 1.1\% (1.0\%) of $I_c$ cannot be taken as significant. Application of the corresponding thresholds removed the chequerboard pattern from the data. The 1\,\%-level needed for the deconvolved G{\"o}ttingen FPI spectra prevents the detection of weak and diffuse magnetic fields. Using the terminology of \citet{stenflo2010} that solar magnetic fields in the quiet Sun exist in either a ``collapsed'' (partly evacuated and concentrated magnetic flux) or an ``uncollapsed'' (diffuse and weak magnetic fields) state, the G{\"o}ttingen FPI can only sense the collapsed fields. This might explain why previous studies of the QS with the G{\"o}ttingen FPI \citep[e.g.,][]{cerdena+etal2003,cerdena+etal2006} found a much larger fraction of  magnetic fields in the kG range in the QS than slit-spectrograph observations of visible or infrared spectral lines \citep[e.g.,][]{khomenko+etal2003,khomenko+etal2005,rezaei+etal2007,orozco+etal2007,gonzalez+etal2008,beck+rezaei2009}. Using a MHD simulation with weak internetwork magnetic fields as a reference, a noise level below about 0.25\% of $I_c$ is needed to detect the weaker half of the polarization signals \citep[for the case of IBIS cf.][]{woeger+etal2009}. This level is not met by neither the G{\"o}ttingen FPI data nor an observation with the Hinode/SP with 4.8\, secs integration time, it requires significantly longer integration times to achieve a sufficiently high S/N ratio. 

The distribution of NCP values for the speckle-deconvolved data matches that of the observations with higher S/N better than the other two data reduction methods but also has seemingly artificially sharp boundaries, a feature that showed up in the NCP maps of all deconvolutions where the information from the NB channel was not or was only partially used. The behavior of the NCP should therefore be related somehow to a step that is specific to these deconvolutions, but neither the MOMFBD weighting both BB and NB channel equally nor the destretching of the spectra. One candidate would be the way of determining and applying the OTF. In the destretching of the spectra, the OTF is not applied directly to the spectra, they are only spatially aligned to the reference BB image. In the MOMFBD weighting both the NB and BB channel equally, the OTF is derived using the full information from the NB channel, whereas in all other cases the information from the BB channel is dominating in the derivation of the OTF. In the presence of (minor) optical differences in the light path of the BB and NB channel, the OTF based on the BB channel might therefore not fit perfectly to the NB channel. The application of the OTF to deconvolve the NB spectra could then result in a small additive effect that only shows up when wavelength-integrated quantities such as the NCP are calculated from the spectra. The NCP maps of deconvolved spectra with a reduced noise level in the NCP -- contrary to their increased rms noise in the Stokes $V$ signal relative to the destretched spectra -- and the overly sharp boundaries in the spatial domain fulfill the definition of showing ``artifacts'' as given above.
\section{Conclusions \label{concl}}
To determine the ``best'' approach for the deconvolution of polarimetric
  spectra, one has to choose a compromise between spatial resolution,
  intensity contrast, and the noise level in the polarization signal. The
  destretched spectra provide a reference for the lowest achievable noise
  level that in our case is roughly preserved by the MOMFBD of the
  spectra with equal weight for BB and NB channel. For the future usage of the G{\"o}ttingen FPI for imaging spectropolarimetry it, however, seems to be necessary to improve the S/N ratio, to be able to lower the final significance thresholds. Especially for weak spectral lines with a small line depth, and hence, small polarization amplitudes, or strong chromospheric lines with a low line-core intensity, the most common polarization amplitudes are far below the 1\,\% level and thus cannot be detected at a low S/N. 
\begin{acknowledgements}
The VTT is operated by the Kiepenheuer-Institut f\"ur Sonnenphysik (KIS) at the
Spanish Observatorio del Teide of the Instituto de Astrof\'{\i}sica de Canarias (IAC). K.G.P. acknowledges financial support by the Spanish Ministry of Science and Innovation through ESP 2006-13030-C06-01, AYA2007-65602, and the European Commission through the SOLAIRE Network (MTRN-CT-2006-035484). K.G.P and C.B. acknowledge partial support by the Spanish Ministry of Science and Innovation through project AYA2010--18029. We thank C. Denker, M. Collados, A. Tritschler, J.A. Bonet, F.~W\"oger and M. L\"ofdahl for fruitful discussions. We thank O. Steiner And R. Rezaei for making the spectra of the MHD simulation available to us.
\end{acknowledgements}
\bibliographystyle{aa}
\bibliography{references_gfpi}
\begin{appendix}
\section{Deconvolution test runs\label{appa}}
\begin{figure*}
\resizebox{17.6cm}{!}{\hspace*{.5cm}\includegraphics{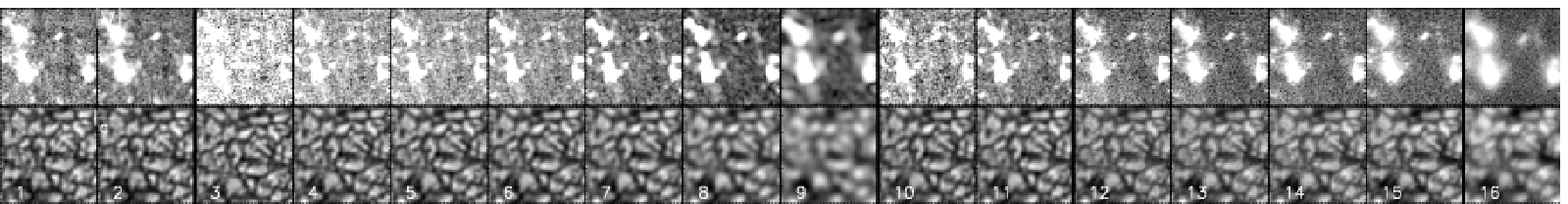}}$ $\\
\caption{Continuum intensity ({\em bottom row}) and average unsigned Stokes $V/I$ signal $< 1.5 \%$ ({\em top row}) for the deconvolution test runs. \label{test_2d}}
\end{figure*}
\begin{table*}
\caption{Intensity contrast of the NB channel and rms noise in $V/I$ in the test runs.\label{test_settings}}
\begin{tabular}{c|cc|ccccccc}\hline\hline
no. & 1 & 2 & 3 & 4 & 5 & 6 & 7 & 8 & 9  \cr\hline
method & Speckle & Speckle & MOMFBD & MOMFBD & MOMFBD & MOMFBD & MOMFBD & MOMFBD & MOMFBD  \cr
filter & -- & Optimum & -- &  0\farcs24&  0\farcs29 &  0\farcs35&  0\farcs45 &  0\farcs63 &  1\farcs05 \cr
weight BB & 1 & 1 & 1 & 1 & 1 & 1 & 1 & 1& 1  \cr
weight NB & 0 & 0 & 0 & 0 & 0 & 0 & 0 & 0& 0  \cr
contrast $[\%]$ & 6.3 & 6.3 & 10.1 & 9.5 & 9.2  & 8.9  & 8.2  & 7.1  & 4.9 \cr
rms V/I $[\%]$ & 0.65 & 0.50 & 1.18 & 0.87 & 0.78 & 0.67 & 0.51 & 0.32 & 0.13 \cr\hline
\end{tabular}\\$ $\\$ $\\
\begin{tabular}{c|cc|cccc|c}\hline\hline
no. & 10 & 11 & 12 & 13 & 14 & 15 & 16  \cr\hline
method & MOMFBD & MOMBFD & MOMFBD & MOMFBD &   MOMFBD &  MOMFBD   & Destr.\cr
weight BB & 1 & 1 & 1 & 1& 1 & 1   & 1     \cr
weight NB & 0 & 0  & 0.1 & 0.25 & 0.5 & 1 & 0  \cr
filter & NF 1.5 & NF 2 & --  & -- &  -- & -- & --\cr
contrast $[\%]$ & 11.1 & 8.5 & 5.8 & 4.9 & 4.9 & 4.3 & 3.6\cr
rms V/I $[\%]$ & 0.89 & 0.59 & 0.54  & 0.41 & 0.41 &  0.32& 0.29\cr\hline
\end{tabular}
\end{table*}
We run a series of deconvolutions with different settings. For the speckle deconvolution, we used on the one hand no Fourier filtering during the deconvolution and on the other hand an optimum filter (no.~1 and 2 in Table \ref{test_settings}). A series of MOMFBDs was done starting from one weighting only the BB channel in the deconvolution (no.~3). The spectra of no.~3 were then Fourier-filtered post facto with a Hamming filter in the Fourier power of individual wavelengths and Stokes parameters (no.~4 to 9). The post-facto filter was set to cut all power beyond a given spatial scale ({\em second row} of Table \ref{test_settings}). Instead of a post-facto filter, we also changed the noise threshold in the MOMFBD already (no.~10 and 11), still only weighting the BB channel. All other MOMFBDs used the default noise filter (NF=1). In the final series, we included the information from both the BB and NB channel with an increasing weight for the NB channel from 0.1 to 1 (no.~12 to 15). The last entry (no.~16 in Table \ref{test_settings}) corresponds to the destretching of the spectra. The spectral noise filter of Sect.~\ref{noise_filter} was not applied, the values of the noise in $V/I$ at continuum wavelengths (630.275\,nm to 630.291\,nm) thus correspond to the {\em first column} of Table \ref{tab_noise}.
\subsection{Reasonable parameter range}
The destretched spectra (no.~16) provide the minimum noise level in $V/I$ (0.29\,\%) and the minimum contrast (3.6\,\%). The maximum contrast is provided by the speckle deconvolution without filter (no.~1, 6.3\,\%) and the MOMFBD deconvolutions weighting only the BB channel without additional filter or with a slightly enhanced noise filter (no.~3 and 10, 10\,--\,11\,\%). The ``best'' settings have to yield contrast values within this range, without a rms noise in $V/I$ {\em below} that of the destretched spectra. The upper limit of the noise level is not restricted ($>$ 1\,\% for no.~3).
\subsection{Contrast {\em vs.} noise}
\begin{figure}
\resizebox{8.8cm}{!}{\includegraphics{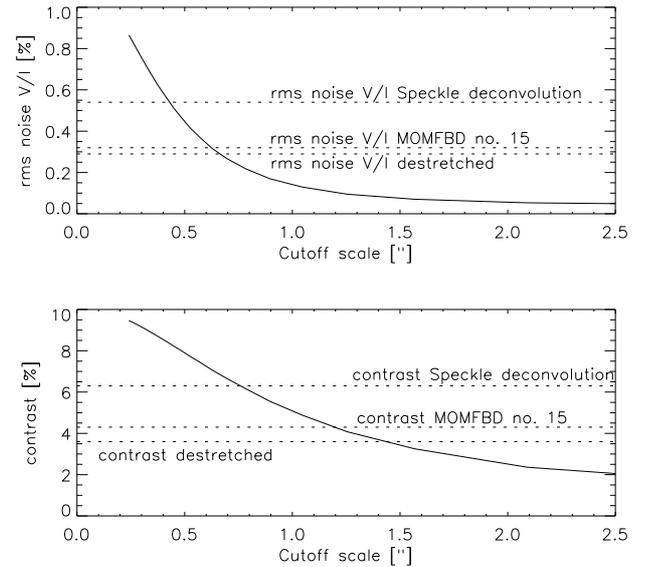}}
\caption{Dependence of intensity contrast ({\em bottom}) and rms noise in $V/I$ ({\em top}) in the MOMFBD spectra on the cutoff scale of a post-facto Fourier filtering. The {\em horizontal dotted} lines give the corresponding values for the speckle-deconvolved, MOMFBD (no.~15), and destretched spectra, respectively. \label{test_filter}}
\end{figure}

Figure \ref{test_2d} shows a subsection of the full FOV in continuum intensity ({\em bottom row}) and unsigned average Stokes $V/I$ signal ({\em top row}) for the different deconvolution settings. The variation of the noise level can be clearly seen. Variations of the intensity contrast are hidden because each panel is scaled individually, only the cases with a degraded spatial resolution (no.~8, 9, and 16) can be distinguished at once. Figure \ref{test_filter} visualizes the trade-off between noise and contrast for the case of the post-facto filtering in the MOMFBD spectra. The noise level falls much more rapidly with the increasing spatial scale at which the power is cut than the intensity contrast. The noise level in the destretched spectra sets an upper limit of about 0\farcs7 for the filter cutoff. The continuum intensity images in Fig.~\ref{test_2d} set a stricter limit. Between no.~7 and no.~8 (0\farcs35 and 0\farcs45), a degradation of the image quality can be clearly seen, whereas for the previous steps (no.~3 to 7) the degradation is not prominent. Larger filter widths significantly smear out the spatial resolution or fall below the noise level of the destretch (no.~9).

All MOMFBDs where the NB channel was not fully weighted give a slightly more diffuse image, which is more prominent in images of the full FOV (not shown here). All MOMFBDs with only a partial weight for the NB channel (no.~12 to 14) had some more or less pronounced artifacts in the spectra, mainly some individual badly reconstructed subfields.
\subsection{Net circular polarization (NCP)\label{ncp_test}}
\begin{figure}
\resizebox{8.8cm}{!}{\hspace*{.5cm}\includegraphics{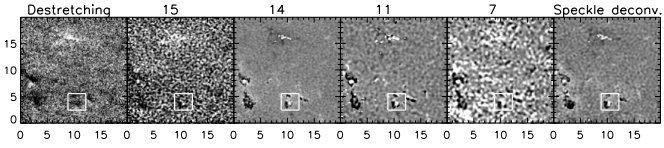}}$ $\\
\caption{NCP maps of some of the test deconvolutions. The display range is $\pm$ 0.15\,pm. The FOV corresponds to the one used in Fig.~\ref{ncp_fig}.\label{ncp_comp_test}}
\end{figure}
The NCP is a sensitive measure to determine the spectral purity of the
deconvolved spectra, in addition to the noise level. The scatter in the NCP
values should increase with the increasing noise level in $V/I$. It, however,
turned out that this is not necessarily the case. Figure \ref{ncp_comp_test}
shows a subsection of the NCP maps of the full FOV for some
deconvolutions. For the methods no.~14 (MOMFBD weighting NB with 0.5), no.~11 (MOMFBD only weighting BB, NF\,=\,2), no.~7 (MOMFBD only weighting BB, post-facto filter cutting at 0\farcs45), and no.~2 (speckle deconvolution) the scatter in the NCP is strongly suppressed and does not scale with the noise level. The NCP shows artifically sharp boundaries in the NCP maps, changing from about zero to the display threshold of $\pm 0.15\,$pm from one pixel to the next, e.g., inside the  {\em white rectangle} marked in Fig.~\ref{ncp_comp_test}. This behavior is especially seen for the deconvolutions where the information of the NB channel was not or only partially used, with the possible exception of no.~7. Neither the destretching nor the MOMFBD weighting both NB and BB channel equally (no.~15) show the sharp boundaries. The latter deconvolution is the only one, where the noise in the NCP of the deconvolved spectra increases with the noise level in $V/I$ (compare the first two maps of Fig.~\ref{ncp_comp_test}). The spectral noise filter of Sect.~\ref{noise_filter} was applied before the calculation of the NCP, but it has no effect on the NCP, and especially does not cause the artifical behavior of the NCP. Data without the spectral filter show the same patterns.
\subsection{Amplification of the polarization amplitude\label{noise_ampl}}
\begin{figure}
\resizebox{8.8cm}{!}{\includegraphics{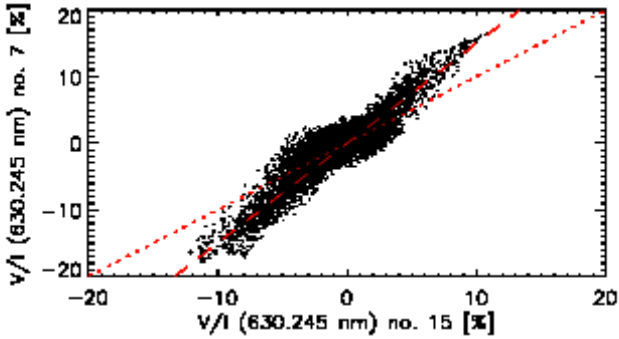}}
\caption{Stokes $V/I\,({\rm \lambda = 630.245\,nm})$ amplitude of deconvolution no.~7 {\em vs.}~that of no.~15. The {\em short-dashed} line denotes a unity slope, the {\em long-dashed} line a slope of 1.5.\label{v_ampl_comp}}
\end{figure}
The deconvolution does not only scale up the noise level with increasing
intensity contrast, but also the amplitude of the polarization signals. Figure
\ref{v_ampl_comp} shows a scatter plot of the polarization amplitudes at
630.245\,nm in method no.~15 (NB fully weighted) and no.~7 (only BB weighted,
post-facto filter cutting at 0\farcs45). The method with the larger contrast
and noise (no.~7) has increased polarization amplitudes, but not on all
pixels. For the largest fraction of pixels inside the FOV, the relation gives
a unity slope between the two co-spatial polarization amplitudes. For all
high-amplitude signals ($|V/I\,({\rm  \lambda = 630.245\,nm})| > 5$\,\%) the slope changes to 1.5, implying a
relative increase of the polarization by 50\,\%, e.g., 7.5\,\% instead of
5\,\%. A direct comparison of the two $V/I\,({\rm  \lambda = 630.245\,nm})$ maps showed that the polarization amplitude in the central part of the network patches is enhanced in no.~7 and reduced in their surroundings, i.e., the ``halo'' produced by spatial smearing was better removed than in no.~15. The increase of the polarization amplitudes is, however, the only argument in favor of the deconvolutions with a contrast
above that of no.~15.
\subsection{Final choice of settings for MOMFBD\label{motiv}}
The results of the various test runs differ mainly in two parameters, the rms
contrast in intensity and the noise level in $V/I$, with a clear trade-off between the two parameters. The spatial resolution seems to be virtually unaffected as long as no strong spatial filtering is involved. Because finally the
polarimetric properties of the spectra are the observed quantities that are
used to derive the properties of the solar atmosphere, we chose the MOMFBD
with the lowest noise level (no.~15) in which NB and BB channel entered with
equal weight in the MOMFBD. Besides the noise level, this choice also yields more reasonable values of the NCP than some of the other deconvolutions. The impact of the choice on derived solar properties can be easily tested by, e.g.,
comparing the results of an inversion of the speckle deconvolved spectra
(medium noise and contrast) with those of an inversion of the spectra of
MOMFBDs no.~3 and 15, and the destretched spectra. The rms contrast should
enter into the spatial variation of the temperature with hopefully a minor
change of the derived magnetic field properties. This is, however, beyond the
scope of the present investigation. 
\end{appendix}
\end{document}